\documentclass[11pt]{article}
\usepackage{latexsym,amssymb,amsmath,times}
\usepackage{enumitem}
\usepackage{float}

\usepackage{hyperref}
\usepackage[square,sort&compress,comma,numbers]{natbib}
\makeatletter

\@addtoreset{equation}{section} \makeatother

\input amssym.def
\input amssym.tex

\usepackage{tikz}
\usetikzlibrary{decorations.markings}

\newcommand{\dynkinradius}{.11cm}
\newcommand{\dynkinstep}{0.7cm}
\newcommand{\dynkindot}[2]{ \fill(\dynkinstep*#1,\dynkinstep*#2) circle (\dynkinradius);}

\newcommand{\dynkinline}[4]{\draw[thick] (\dynkinstep*#1,\dynkinstep*#2) -- (\dynkinstep*#3,\dynkinstep*#4);}
\newcommand{\dynkindots}[4]{\draw[thick,dotted] (\dynkinstep*#1,\dynkinstep*#2) -- (\dynkinstep*#3,\dynkinstep*#4);}

\newenvironment{dynkin}{\begin{tikzpicture}[decoration={markings,mark=at position 0.7 with {\arrow{>}}}]}
{\end{tikzpicture}}

\newcommand{\half}{{{\textstyle\frac{1}{2}}}}
\newcommand{\quarter}{{{\textstyle\frac{1}{4}}}}
\newcommand{\be}{\begin{equation}}
\newcommand{\ee}{\end{equation} }
\newcommand{\beqa}{\begin{eqnarray} }
\newcommand{\eeqa}{\end{eqnarray} }
\newcommand{\ba}{\begin{array}}
\newcommand{\ea}{\end{array}}

\newcommand{\Spin}{\mathbf{Spin}}

\newcommand{\rmd}{{\rm d}}

\newcommand{\rmG}{{\rm G}}

\newcommand{\bpr}{b^{\prime}}
\newcommand{\cpr}{c^{\prime}}

\newcommand\hcL{{\hat{\cal L}}}

\newcommand{\ODD}{\mathbf{O}(D,D)}

\newcommand{\Ott}{\mathbf{O}(10,10)}

\newcommand{\SLf}{\mathbf{SL}(5)}

\newcommand{\SLN}{\mathbf{SL}(N)}

\newcommand{\slN}{\mathbf{sl}(N)}

\newcommand\cK{{\cal K}}

\newcommand\cM{{\cal M}}
\newcommand\cN{{\cal N}}

\newcommand\cP{{\cal P}}

\newcommand\cR{{\cal R}}

\newcommand\cX{{\cal X}}
\newcommand\cY{{\cal Y}}

\newcommand\hA{\hat{A}}

\newcommand\dis{\displaystyle}

\newcommand\seceq{=}

\def\tg{\tilde{g}}

\def\tv{\tilde{v}}

\def\tpartial{\tilde{\partial}}

\def\tphi{\tilde{\phi}}

\newcommand{\na}{{\nabla}}
\newcommand{\trd}{{\bigtriangledown}}

\begin{document}
\begin{titlepage}
\title{
\vskip 2cm   U-gravity\,: $\SLN$\\
~}
\author{\sc Jeong-Hyuck Park\quad and\quad Yoonji Suh\quad\quad}
\date{}
\maketitle \vspace{-1.0cm}
\begin{center}
\texttt{~~~~~ park@sogang.ac.kr\quad\quad\quad\quad yjsuh@sogang.ac.kr\quad}\\
~\\
Department of Physics, Sogang University, Mapo-gu,  Seoul 121-742, Korea\\
~{}\\
~~~\\~\\
\end{center}
\begin{abstract}
\vskip0.2cm
\noindent 
We construct a duality manifest gravitational theory for the   special linear group,  ${\mathbf{SL}(N)}$ with $N{\neq 4}$.  The spacetime is  formally extended, to have the dimension   $\textstyle{\frac{1}{2}} N(N-1)$, yet  is   \textit{gauged}.  Consequently      the theory is subject to a section condition.  We introduce   a semi-covariant derivative and  a semi-covariant  `Riemann' curvature,  both of which  can be   completely  covariantized   after symmetrizing or contracting     the ${\mathbf{SL}(N)}$ vector indices properly.  Fully covariant scalar and `Ricci' curvatures then constitute  the  action and the `Einstein'   equation of motion.  For $N\geq 5$, the section  condition  admits   duality  inequivalent two  solutions, one $(N-1)$-dimensional and the other three-dimensional.  In each case, the theory can describe not only Riemannian but also  non-Riemannian    backgrounds. 
\end{abstract}


\thispagestyle{empty}

\end{titlepage}
\newpage
\tableofcontents 

\section{Introduction}
While Lorentz symmetry unifies    space and time into spacetime,  duality requires further  extension  of the spacetime~\cite{Cremmer:1979up,Duff:1989tf,Duff:1990hn}. T-duality in string theory becomes  a manifest   $\ODD$ rotation in doubled spacetime~\cite{Duff:1989tf,Duff:1990hn,Tseytlin:1990nb,Tseytlin:1990va,Siegel:1993xq,Siegel:1993th,Grana:2008yw,Coimbra:2011nw,Coimbra:2012yy},  and so do various $\cM$-theory U-dualities in extended spacetimes, including the maximal $E_{11}$~\cite{West:2001as,West:2004st,Bergshoeff:2007qi,Bergshoeff:2007vb,West:2010ev,Rocen:2010bk,West:2011mm}, $E_{10}$~\cite{Damour:2002cu,Nicolai:2003fw,Kleinschmidt:2004dy,Damour:2007dt} and smaller cousins~\cite{Berman:2010is,Berman:2011cg,Berman:2011pe,Berman:2011jh,Thompson:2011uw,Coimbra:2011ky,Berman:2012vc,Godazgar:2013rja,
Park:2013gaj,Aldazabal:2013mya,Cederwall:2013naa,Cederwall:2013oaa,Hohm:2013jma,Hohm:2013pua,Hohm:2013vpa,Hohm:2013uia,Aldazabal:2013via,Godazgar:2014sla,Malek:2012pw} (see also \cite{Aldazabal:2013sca,Berman:2013eva,Hohm:2013bwa} for  further references).\\

\noindent Unlike the Lorentz symmetry unification of  space and  time, the duality manifest extension of the spacetime  calls for the existence of seemingly unphysical `dual' spacetime.  One simple prescription to eliminate  this unphysical feature  is to let all the fields be independent of the dual coordinates, \textit{e.g.~}\cite{Siegel:1993xq,Siegel:1993th} and `Generalized Geometry'~\cite{Gualtieri:2003dx,Hitchin:2004ut,Hitchin:2010qz,Grana:2008yw,Pacheco:2008ps,Koerber:2010bx,Coimbra:2011nw,Coimbra:2012yy}.  More covariant method  is to enforce  so called a \textit{section condition}  on all the functions defined on the extended spacetime. The section condition is a differential constraint and can be solved by a certain hyper-subspace, called `\textit{section}',   on which the theory is restricted to live. Duality then rotates the section in the extended spacetime. Especially, acting  on an isometry direction,   it  may produce  a new solution while the section can  remain    unrotated.  This is the very  geometric insight that has motivated  \cite{Siegel:1993xq,Siegel:1993th} or   Double Field Theory (DFT)~\cite{Hull:2009mi,Hull:2009zb,Hohm:2010jy,Hohm:2010pp}. Fixing the section explicitly and parametrizing the DFT variables by Riemannian ones, DFT may locally reduce   to Generalized Geometry.   Then, like   T-fold~\cite{Hull:2004in,Hull:2006va,Hull:2006qs}, by combining diffeomorphism and $\ODD$ rotation as for a transition function,   DFT may acquire  nontrivial  global aspects of non-geometry~\cite{Hohm:2012gk,Park:2013mpa,Berman:2014jba,Cederwall:2014kxa,Papadopoulos:2014mxa}. \\

\noindent Further,  once formulated in terms of genuine $\ODD$ covariant variables, DFT   does not  merely  repackage  Generalized Geometry or known supergravities, but  can also describe  non-Riemannian  backgrounds where  the notion of Riemannian metric ceases to exist even \textit{locally}~\cite{Lee:2013hma}.  In a somewhat abstract level, the DFT-metric can be defined simply as a `symmetric $\ODD$ element', with which  (bosonic) DFT and  a doubled string world-sheet  action~\cite{Lee:2013hma} still make sense.   For most (``non-degenerate") cases  the DFT-metric  can be parametrized by Riemannian metric, $g_{\mu\nu}$  and Kalb-Ramond $B$-field, which allows  DFT to describe an ordinary Riemannian gravity.   But, for the other  (``degenerate")  cases  the DFT-metric may not admit any Riemannian interpretation, even locally! An extreme   example is  the DFT vacuum solution  where    the DFT-metric coincides with the $\ODD$ invariant constant metric. The doubled string  action then  reduces to  a chiral sigma model~\cite{Lee:2013hma}, similar to \cite{Gomis:2000bd}.\\

\noindent   As demonstrated  in Refs.\cite{Park:2013mpa,Lee:2013hma}, imposing the section condition  is, in fact,  equivalent  to   postulating an equivalence relation on the doubled coordinate space. That is to say, \textit{spacetime is doubled yet gauged}.  Accordingly, each equivalence class or gauge orbit   represents a single physical point, and  diffeomorphism symmetry means an invariance under arbitrary reparametrizations of the gauge orbits. This  allows  more than one   finite  transformation rule of diffeomorphism~\cite{Hohm:2012gk,Park:2013mpa,Berman:2014jba}.  The idea has been  pushed further to construct a completely covariant  string world-sheet action on doubled-yet-gauged spacetime~\cite{Lee:2013hma}, where the coordinate gauge symmetry is    realized literally as one of the   local  symmetries  of the action.   In a way,  understanding    the section condition by    gauged  spacetime   agrees with the lesson learned  from the 20th century that `local symmetry dictates fundamental physics'.   \\
~\\

\begin{table}[H]
\begin{center}
\begin{tabular}{ll}
\hline

\mbox{~~~\quad}\begin{dynkin}
   \dynkindot{1}{0}
   \dynkindot{2}{0}
   \dynkindot{3}{0}
   \dynkindots{4}{0}{6}{0}
    \dynkindot{7}{0}
    \dynkindot{8}{0}
    \dynkindot{9}{0}
    \dynkindot{10}{0}
    \dynkinline{1}{0}{4}{0}
    \dynkinline{6}{0}{8}{0}
  
    \dynkinline{8}{0}{10}{0}
  \end{dynkin}~&~ $A_{N-1}\equiv\,\slN$~~~~\\{}&{}\\
  
  \mbox{~~~\quad}\begin{dynkin}
   \dynkindot{1}{0}
   \dynkindot{2}{0}
   \dynkindot{3}{0}
   \dynkindots{4}{0}{6}{0}
    \dynkindot{7}{0}
    \dynkindot{8}{0}
    \dynkindot{8}{1}
    \dynkindot{9}{0}
  
    \dynkinline{1}{0}{4}{0}
    \dynkinline{6}{0}{8}{0}
  \dynkinline{8}{0}{8}{1}
    \dynkinline{8}{0}{9}{0}
  \end{dynkin}~&~ $D_{N-1}\equiv\,\mathbf{so}(N{-1},N{-1})$~~~~\\{}&{}\\

\mbox{\qquad\qquad\!\!\!}\begin{dynkin}
 
   \dynkindot{2}{0}
   \dynkindot{3}{0}
   \dynkindots{4}{0}{6}{0}
    \dynkindot{7}{0}
    \dynkindot{8}{0}
    \dynkindot{8}{1}
    \dynkindot{9}{0}
    \dynkindot{10}{0}
    \dynkinline{2}{0}{4}{0}
    \dynkinline{6}{0}{8}{0}
    \dynkinline{8}{0}{8}{1}
    \dynkinline{8}{0}{10}{0}
  \end{dynkin}\,~&~ $E_{N-1}$~~~~\\{}&{}\\

\mbox{~~~\quad}\begin{dynkin}
   \dynkindot{1}{0}
   \dynkindot{2}{0}
   \dynkindot{3}{0}
   \dynkindots{4}{0}{6}{0}
    \dynkindot{7}{0}
    \dynkindot{8}{0}
    \dynkindot{8}{1}
    \dynkindot{9}{0}
    \dynkindot{10}{0}
    \dynkinline{1}{0}{4}{0}
    \dynkinline{6}{0}{8}{0}
    \dynkinline{8}{0}{8}{1}
    \dynkinline{8}{0}{10}{0}
  \end{dynkin}~&~ $E_{N}$~~~~\\
  \hline
\end{tabular}
\caption{Dynkin diagrams  for $A_{N-1}$, $D_{N-1}$, $E_{N-1}$  ~and~  $E_{N}$.} 
\label{TableDynkin}
 \end{center}
\end{table}

\noindent In this note,    we construct  a duality manifest gravitational theory for the   special linear group,  ${\mathbf{SL}(N)}$ with   $N\neq 4$, ,  in  the name\footnote{U-duality manifest theory  has been occasionally  dubbed,  `Exceptional Geometry' or  `Exceptional Field Theory' (EFT), \textit{e.g.~}\cite{Cederwall:2013naa,Cederwall:2013oaa,Hohm:2013jma,Hohm:2013pua,Hohm:2013vpa,Hohm:2013uia,Aldazabal:2013via,Godazgar:2014sla}. However, strictly speaking,  this naming should be  proper  only for  the exceptional groups.  Since our duality group, $\SLN$, is    not exceptional,  we   call  our theory differently.  U-gravity  manifests U-duality and provide a Universal framework for $(N{-1})$-dimensional  and three-dimensional gravities, as well as Riemannian and non-Riemannian geometries.  } of\, `\textbf{$\SLN$ {U-gravity}}'.  The existence of such an $A_{N-1}\equiv\slN$  manifest  geometry has been predicted in  \cite{Baraglia:2011dg,Strickland-Constable:2013xta}  based on   Dynkin diagram analyses.   Thus, our construction  provides an explicit realization of the prediction.    Further, as seen from  Table~\ref{TableDynkin}, $E_{N}$ algebra  contains  three inequivalent  ``maximal" sub-algebras,   $A_{N-1}$, $D_{N-1}$  and $E_{N-1}$. This   implies that there are   three distinct ways of  reducing  the grand   scheme  of $E_{11}$~\cite{West:2001as,West:2004st,Bergshoeff:2007qi,West:2010ev,Rocen:2010bk,West:2011mm}:  \textit{(i)\,} $\mathbf{SL}(11)$ U-gravity,   ~\textit{(ii)\,}  $\mathbf{O}(10,10)$ DFT\,  and   ~\textit{(iii)\,} $E_{10}$ program~\cite{Damour:2002cu,Nicolai:2003fw,Kleinschmidt:2004dy,Damour:2007dt}.\footnote{From \cite{Baraglia:2011dg,Strickland-Constable:2013xta}, any sub-algebra of $E_{11}$ should allow its own generalized geometry. In this context,  it has been shown recently   that the `full'  eleven-dimensional supergravity can be reformulated to manifest  $E_{6}$, $E_{7}$ or $E_{8}$ `sub'-algebras~\cite{Hohm:2013jma,Hohm:2013pua,Hohm:2013vpa,Hohm:2013uia}.  }

\newpage

\noindent While the motivation of our work comes from the  Dynkin diagram  prediction~\cite{Baraglia:2011dg,Strickland-Constable:2013xta} and the $E_{11}$ proposal~\cite{West:2001as,West:2004st,Bergshoeff:2007qi,West:2010ev,Rocen:2010bk,West:2011mm},   for the actual  construction of the $\SLN$ U-gravity, we heavily  employ    the differential geometry tools   from  \cite{Jeon:2010rw,Jeon:2011cn,Park:2013mpa,Lee:2013hma} which were developed to   provide  an underlying  `stringy'   differential geometry for DFT (see Table~\ref{TableFeatures} for its characteristic).\footnote{Alternative approaches include \cite{Hohm:2010xe,Berman:2013uda,Hohm:2011zr,Hohm:2011dv,Hohm:2012mf,Hohm:2011si,Hohm:2011nu}.}    The methods have been successfully applied  to construct    Yang-Mills DFT~\cite{Jeon:2011kp},  coupling to fermions~\cite{Jeon:2011vx}, coupling to  RR-sector~\cite{Jeon:2012kd},  $\cN=1, D=10$ (full order) super DFT~\cite{Jeon:2011sq},   $\cN=2, D=10$  (full order) super DFT~\cite{Jeon:2012hp},  and   the  completely covariant  string action on doubled-yet-gauged spacetime~\cite{Lee:2013hma}.  \\
~\\

\begin{table}[H]
\begin{center}
\begin{tabular}{l}
\hline

~~$\bullet$~~  \textit{Extended-yet-gauged spacetime (section condition).}\\
~~$\bullet$~~  \textit{Diffeomorphism generated by a generalized Lie derivative, \textit{c.f.~}}\cite{Berman:2011cg}.~~\\
~~$\bullet$~~  \textit{Semi-covariant derivative and semi-covariant  Riemann curvature.}\\
~~$\bullet$~~  \textit{Complete  covariantizations  of them dictated  by   a  projection operator.~~}\\

\hline
\end{tabular}
\caption{The common  features of $\SLN$ U-gravity and DFT-geometry~\cite{Jeon:2010rw,Jeon:2011cn,Park:2013mpa,Lee:2013hma}.} 
\label{TableFeatures}
\end{center}
\end{table}
~\\
Especially when $N{=5}$, the constructed  theory of U-gravity  reduces to our preceding research  of  `$\SLf$ U-geometry'~\cite{Park:2013gaj} (\textit{c.f.~}\cite{Berman:2010is}). The present paper  generalizes our previous work   to  an arbitrary special linear group, $\SLN$ with $N\neq 4$, and also contains some  novel findings,  such as      the semi-covariant Riemann curvature and  an eight-index  projection operator.    \\

\noindent  In the next section we spell out all the essential elements  that constitute   the   $\SLN$ U-gravity.  Exposition  will  follow   in  section~\ref{SECFORM}. We conclude with    outlook in the final section.\\

\newpage

\section{Constitution of $\SLN$ U-gravity\label{SECSUMMARY}}
Essential elements that constitute     $\SLN$ U-gravity are as follows.
\begin{itemize}
\item  \textit{{\textbf{Notation}}}.   Small  Latin alphabet letters denote  the $\SLN$ vector indices, \textit{i.e.}  $a,b,c,\cdots=1,2,\cdots,N$.

\item \textit{{\textbf{Extended-yet-gauged spacetime}}}. The spacetime is  formally extended, being  $\half N(N-1)$-dimensional.  The coordinates    carry   a pair of  anti-symmetric $\SLN$  vector indices,
\be
x^{ab}=-x^{ba}=x^{[ab]}\,,
\ee
and hence so does  the  derivative,
\be
\ba{ll}
\partial_{ab}=-\partial_{ba}=\partial_{[ab]}=\frac{\partial~~}{\partial x^{ab}}\,,
\quad&\quad
\partial_{ab}x^{cd}=\delta_{a}^{~c}\delta_{b}^{~d}-\delta_{a}^{~d}\delta_{b}^{~c}\,.
\label{fparderiv}
\ea
\ee
However, \underline{the  spacetime  is gauged}:  the coordinate space is equipped with   an   {equivalence relation}, 
\be
x^{ab}~\sim~x^{ab}+\textstyle{\frac{1}{(N-4)!}}\epsilon^{abc_{1}\cdots c_{N-4}de}\phi_{c_{1}\cdots c_{N-4}}\partial_{de}\varphi\,,
\label{fCGS}
\ee
which we call `coordinate gauge symmetry' (\textit{c.f.} \cite{Park:2013mpa,Lee:2013hma} for  DFT analogy). In (\ref{fCGS}),  $\phi_{c_{1}\cdots c_{N-4}}$ and $\varphi$ are arbitrary --but not necessarily covariant-- functions in  the theory of  U-gravity.  As usual,  $\epsilon^{c_{1}c_{2}\cdots c_{N}}$ denotes the totally anti-symmetric  Levi-Civita symbol with $\epsilon^{12\cdots N}\equiv1$.  Apparently, the above equivalence relation  makes sense for $N\geq 5$. For $N=2,3$, the spacetime is not to be  gauged. 

Each equivalence class, or gauge orbit defined by the equivalence  relation~(\ref{fCGS}),  represents a single physical point, and  diffeomorphism symmetry means  an invariance under arbitrary reparametrizations of the gauge orbits.

\item \textit{{\textbf{Realization of the coordinate gauge symmetry.}}} The equivalence relation  (\ref{fCGS}) is realized  in U-gravity by enforcing  that,   arbitrary  functions and   their arbitrary  derivatives, denoted here collectively  by $\Phi$,    are  invariant under the coordinate gauge symmetry  \textit{shift},  
\be
\ba{ll}
\Phi(x+\Delta)=\Phi(x)\,,\quad&\quad\Delta^{ab}=\textstyle{\frac{1}{(N-4)!}}\epsilon^{abc_{1}\cdots c_{N-4}de}\phi_{c_{1}\cdots c_{N-4}}\partial_{de}\varphi\,.
\ea
\label{fTensorCGS}
\ee
\newpage

\item \textit{{\textbf{Section condition.}}} The invariance under the coordinate gauge symmetry~(\ref{fTensorCGS}) is, in fact,  equivalent to a {section condition},
\be
\partial_{[ab}\partial_{cd]}\equiv 0\,.
\label{fseccon}
\ee
Acting on arbitrary functions, $\Phi$, $\Phi^{\prime}$,  and their products, the section condition leads to 
\begin{eqnarray}
&\partial_{[ab}\partial_{cd]}\Phi=\partial_{[ab}\partial_{c]d}\Phi\seceq 0&\quad(\mbox{weak~~constraint})\,,\label{fseccon1}\\
&\partial_{[ab}\Phi\partial_{cd]}\Phi^{\prime}=
\half\partial_{[ab}\Phi\partial_{c]d}\Phi^{\prime}-\half\partial_{d[a}\Phi\partial_{bc]}\Phi^{\prime}\seceq 0&\quad(\mbox{strong~~constraint})\,.
\label{fseccon2}
\end{eqnarray}

\item \textit{{\textbf{Diffeomorphism.}}}  Diffeomorphism symmetry in $\SLN$ U-gravity    is generated by a generalized Lie derivative, 
\be
\ba{ll}
\hcL_{X}T^{a_{1}a_{2}\cdots a_{p}}{}_{b_{1}b_{2}\cdots b_{q}}:=\!\!&
\half X^{cd}\partial_{cd}T^{a_{1}a_{2}\cdots a_{p}}{}_{b_{1}b_{2}\cdots b_{q}}
+\half(\half p- \half q+\omega)\partial_{cd}X^{cd}T^{a_{1}a_{2}\cdots a_{p}}{}_{b_{1}b_{2}\cdots b_{q}}\\
{}&-\sum_{i=1}^{p}T^{a_{1}\cdots c\cdots a_{p}}{}_{b_{1}b_{2}\cdots b_{q}}\partial_{cd}X^{a_{i}d}
+\sum_{j=1}^{q}\partial_{b_{j}d}X^{cd}T^{a_{1}a_{2}\cdots a_{p}}{}_{b_{1}\cdots c\cdots b_{q}}\,.
\ea
\label{fgLNw}
\ee
Here  we let  the tensor density, $T^{a_{1}a_{2}\cdots a_{p}}{}_{b_{1}b_{2}\cdots b_{q}}$,  carry  the  {`total' weight},  $\half p- \half q+\omega$, such that each upper or  lower  index contributes to the total weight by $+\frac{1}{2}$ or $-\frac{1}{2}$ respectively, while $\omega$ corresponds to a possible  `extra' weight.   

In particular, the generalized Lie derivative  of the  Kronecker delta symbol is   trivial,
\be
\hcL_{X}\delta^{a}_{~b}=0\,,
\ee
and the commutator of the generalized Lie derivatives is closed  by a generalized bracket~\cite{Berman:2011cg},
\be
\ba{ll}
\left[\hcL_{X},\hcL_{Y}\right]=\hcL_{[X,Y]_{\rmG}}\,,\quad&\quad
[X,Y]^{ab}_{\rmG}=\half X^{cd}\partial_{cd}Y^{ab}-\textstyle{\frac{3}{2}}X^{[ab}\partial_{cd}Y^{cd]}\,-\,\left(X\,\leftrightarrow\,Y\right)\,.
\ea
\label{fgB2}
\ee

It is a somewhat   surprising   result   of us that   the  above definition of the generalized Lie derivative --including the  total weight--   is independent of  the rank of the duality group, or $N$, and thus  is  identical to the known  one  in \cite{Berman:2011cg,Coimbra:2012af} for the  case of $N=5$.

\item \textit{{\textbf{U-metric.}}} The only   geometric object in $\SLN$ U-gravity     is  a  metric,  or \textit{U-metric},  which is  a generic    non-degenerate $N\times N$ symmetric matrix, obeying surely the section condition,
\be
M_{ab}=M_{ba}=M_{(ab)}\,.
\label{fMetric}
\ee
Like in Riemannian   geometry, the U-metric with its inverse, $M^{ab}$,  may    freely  lower or raise  the positions of the $N$-dimensional  $\SLN$ vector  indices.

\item \textit{{\textbf{Integral measure.}}}  While the U-metric has no extra weight, its determinant, $M\equiv\det(M_{ab})$, acquires an extra weight, $\omega=4-N$. The duality invariant integral measure  is then
\be
|M|^{\frac{1}{4-N}}\,.
\label{fmeasure}
\ee

\item  \textit{{\textbf{Semi-covariant derivative and semi-covariant  Riemann curvature.}}} We define    a semi-covariant derivative, 
\be
\ba{ll}
\na_{cd}T^{a_{1}a_{2}\cdots a_{p}}{}_{b_{1}b_{2}\cdots b_{q}}:=&
\partial_{cd}T^{a_{1}a_{2}\cdots a_{p}}{}_{b_{1}b_{2}\cdots b_{q}}
+\half(\half p- \half q+\omega)\Gamma_{cde}{}^{e}T^{a_{1}a_{2}\cdots a_{p}}{}_{b_{1}b_{2}\cdots b_{q}}\\
{}&~-\sum_{i=1}^{p}T^{a_{1}\cdots e\cdots a_{p}}{}_{b_{1}b_{2}\cdots b_{q}}\Gamma_{cde}{}^{a_{i}}+\sum_{j=1}^{q}\Gamma_{cdb_{j}}{}^{e}T^{a_{1}a_{2}\cdots a_{p}}{}_{b_{1}\cdots e\cdots b_{q}}\,,
\ea
\label{fsemicov}
\ee
and a semi-covariant  Riemann curvature,
\be
\ba{rl}
S_{abcd}:=&3\partial_{[ab}\Gamma_{e][cd]}{}^{e}+3\partial_{[cd}\Gamma_{e][ab]}{}^{e}+\quarter\Gamma_{abe}{}^{e}\Gamma_{cdf}{}^{f}+\half\Gamma_{abe}{}^{f}\Gamma_{cdf}{}^{e}\\
{}&+\Gamma_{ab[c}{}^{e}\Gamma_{d]ef}{}^{f}+\Gamma_{cd[a}{}^{e}\Gamma_{b]ef}{}^{f}+\Gamma_{ea[c}{}^{f}\Gamma_{d]fb}{}^{e}-\Gamma_{eb[c}{}^{f}\Gamma_{d]fa}{}^{e}\,.
\ea
\label{fsemicovS}
\ee
The semi-covariant derivative obeys  the Leibniz rule and annihilates the Kronecker delta symbol, 
\be
\na_{cd}\delta^{a}_{~b}=0\,.
\ee
A crucial defining property of the semi-covariant  Riemann curvature is that, under arbitrary transformation of the connection it   transforms as  \textit{total derivative},
\be
\delta S_{abcd}=3\na_{[ab}\delta\Gamma_{e][cd]}{}^{e}+
3\na_{[cd}\delta\Gamma_{e][ab]}{}^{e}\,.
\label{fdeltaS4}
\ee
Further, the semi-covariant  Riemann curvature   satisfies precisely the same symmetric properties as the ordinary  Riemann curvature, including the Bianchi identity,  
\be
\ba{ll}
S_{abcd}=S_{[ab][cd]}=S_{cdab}\,,\quad&\quad S_{[abc]d}=0\,.
\ea
\ee

\item \textit{{\textbf{Connection.}}} The connection of   the semi-covariant derivative and the semi-covariant   Riemann  curvature    is given by
\be
\ba{ll}
\Gamma_{abcd}&=A_{abcd}+\half(A_{acbd}-A_{adbc}+A_{bdac}-A_{bcad})\\
{}&\quad+\textstyle{\frac{1}{N-2}}\left(M_{ac}A^{e}{}_{(bd)e}-M_{ad}A^{e}{}_{(bc)e}+M_{bd}A^{e}{}_{(ac)e}-M_{bc}A^{e}{}_{(ad)e}\right)\,,
\ea
\label{fGamma}
\ee
where we set
\be
A_{abcd}:=-\half\partial_{ab}M_{cd}+\textstyle{\frac{1}{2(N-4)}}M_{cd}\partial_{ab}\ln\left|M\right|\,.
\label{fdefA}
\ee
This connection  is  \textit{the unique solution} to the following five constraints:\footnote{See  \cite{Jeon:2010rw,Jeon:2011cn} for  the analogous   constraints in DFT.}
\begin{eqnarray}
&&\Gamma_{abcd}+\Gamma_{abdc}=2A_{abcd}\,,\label{fcompM}\\
&&\Gamma_{abc}{}^{d}+\Gamma_{bac}{}^{d}=0\,,\label{fGab}\\
&&\Gamma_{abc}{}^{d}+\Gamma_{bca}{}^{d}+\Gamma_{cab}{}^{d}=0\,,\label{fGabc}\\
&&\Gamma_{cab}{}^{c}+\Gamma_{cba}{}^{c}=0\,,\label{fGcc}\\
&&\cP_{abcd}{}^{efgh}\Gamma_{efgh}=0\,.\label{flastcon}
\end{eqnarray}
The first relation~(\ref{fcompM}) is equivalent to the   U-metric compatibility condition,
\be
\na_{ab}M_{cd}=0\,.
\ee
The second condition~(\ref{fGab}) is natural from the skew-symmetric nature of the coordinates,  $x^{(ab)}=0$ and hence  $\partial_{(ab)}=\na_{(ab)}=0$.  The next   two constraints, (\ref{fGabc}) and (\ref{fGcc}), make the semi-covariant derivative   compatible  with  the generalized Lie derivative as well as with  the generalized bracket, 
\be
\ba{ll}
\hcL_{X}(\partial)=\hcL_{X}(\na)\,,\quad&\quad
[X,Y]_{\rmG}(\partial)=[X,Y]_{\rmG}(\na)\,.
\ea
\ee  
The last  formula~(\ref{flastcon}) is a projection condition  which   we  impose intentionally      in order to ensure the uniqueness.  

\item \textit{{\textbf{Projection operator.}}}  The eight-index  projection operator, used in (\ref{flastcon}),  is explicitly,
\be
\ba{ll}
\cP_{abcd}{}^{klmn}=&
\half\delta_{[a}^{~[k}\delta_{b]}^{~l]}\delta_{[c}^{~[m}\delta_{d]}^{~n]}
+\half\delta_{[c}^{~[k}\delta_{d]}^{~l]}\delta_{[a}^{~[m}\delta_{b]}^{~n]}
+\half M_{c[a}\delta_{b]}^{~m}M^{n[k}\delta_{d}^{~l]}
-\half M_{c[a}\delta_{b]}^{~[k}M^{l]n}\delta_{d}^{~m}\\
{}&~+\textstyle{\frac{1}{N-2}}\left(
\delta_{[a}^{~n}M_{b][c}M^{m[k}\delta_{d]}^{~l]}
+\delta_{[c}^{~n}M_{d][a}M^{m[k}\delta_{b]}^{~l]}
-M_{c[a}M_{b]d}M^{m[k}M^{l]n}\right)\,,
\ea
\label{fprojection}
\ee
which  satisfies
\be
\ba{ll}
\cP_{abcd}{}^{pqrs}\cP_{pqrs}{}^{klmn}=\cP_{abcd}{}^{klmn}\,,\quad&\quad \cP_{abs}{}^{sklmn}=0\,,\\
\cP_{abcd}{}^{klmn}=\cP_{[ab]cd}{}^{[kl]mn}\,,\quad&\quad
\cP_{ab[cd]}{}^{klmn}=\cP_{cd[ab]}{}^{klmn}\,.
\ea
\ee
Crucially, the projection operator  dictates    the  anomalous terms   in the  diffeomorphic transformations    of   the semi-covariant derivative and  the semi-covariant  Riemann curvature,
\be
\ba{rl}
\!\!\!(\delta_{X}-\hcL_{X})(\na_{ab}T^{c_{1}\cdots c_{p}}{}_{d_{1}d_{2}\cdots d_{q}})&=-\sum_{i=1}^{p}T^{c_{1}\cdots e\cdots c_{p}}{}_{d_{1}\cdots d_{q}}\Omega_{abe}{}^{c_{i}}
+\sum_{j=1}^{q}\Omega_{abd_{j}}{}^{e}T^{c_{1}c_{2}\cdots c_{p}}{}_{d_{1}\cdots e\cdots d_{q}}\,,\\
(\delta_{X}-\hcL_{X})S_{abcd}&=2\na_{e[a}\Omega_{b][cd]}{}^{e}
+2\na_{e[c}\Omega_{d][ab]}{}^{e}\,,\\
\Omega_{abcd}&=\cP_{abcd}{}^{klm}{}_{n}\partial_{kl}\partial_{me}X^{ne}\,.
\ea
\ee

\item \textit{{\textbf{Complete  covariantizations. }}} Both the  semi-covariant derivative and  the semi-covariant  Riemann curvature can be fully covariantized    by  (anti-)symmetrizing or contracting  the $\SLN$ vector indices properly~\cite{Park:2013gaj},
\be
\ba{llll}
\na_{[ab}T_{c_{1}c_{2}\cdots c_{q}]}\,,~~~&~~~\na_{ab}T^{a}\,,~~~&~~~
\na^{a}{}_{b}T_{[ca]}+\na^{a}{}_{c}T_{[ba]}\,,~~~&~~~\na^{a}{}_{b}T_{(ca)}-\na^{a}{}_{c}T_{(ba)}\,,\\
\multicolumn{4}{c}{\na_{ab}T^{[abc_{1}c_{2}\cdots c_{q}]}~~~(\mbox{divergence})\,,~~~~~~~~
\na_{ab}\na^{[ab}T^{c_{1}c_{2}\cdots c_{q}]}~~~(\mbox{Laplacian})\,,}
\ea
\label{fCOMPLETECOV}
\ee
and
\be
\ba{ll}
S_{ab}:=S_{acb}{}^{c}=S_{ba}~~&\quad(\mbox{Ricci~curvature})\,,\\
S:=M^{ab}S_{ab}=S_{ab}{}^{ab}~~&\quad(\mbox{scalar~curvature})\,.
\ea
\label{fRicciScalar}
\ee

\item  \textit{{\textbf{Action.}}} The action of  $\SLN$ U-gravity     is given by the fully covariant scalar curvature, 
\be
\dis{\int_{\Sigma}}M^{\frac{1}{4-N}\,}S\,,
\label{factionR}
\ee
where the integral is taken over a section, $\Sigma$. 

\item \textit{{\textbf{The Einstein equation of motion.}}} The  equation of motion corresponds to the vanishing of the `Einstein' tensor,
\be
S_{ab}+ \textstyle{\frac{1}{2(N-4)}}M_{ab}S=0\,.
\label{fEOM}
\ee
Diffeomorphism symmetry of the action implies a conservation relation, 
\be
\na^{c}{}_{[a}S_{b]c}+\textstyle{\frac{3}{8}}\na_{ab}S=0\,.
\label{fconservation}
\ee 

\item \textit{{\textbf{Two inequivalent   sections.}}}   Up to $\SLN$ duality rotations, there exist  two inequivalent  solutions to the section condition, which we denote here by $\Sigma_{N-1}$ and $\Sigma_{3}$.
\begin{enumerate}
\item $\Sigma_{N-1}$ is an $(N-1)$-dimensional section given by
\be
\ba{ll}
\partial_{\alpha\beta} = 0\,, \quad &\quad 
\partial_{\alpha N}\neq 0 \,,
\ea
\label{fMsection}
\ee
where $\alpha, \beta = 1,2, \cdots, N-1$. 

\item $\Sigma_{3}$ is a three-dimensional section characterized  by
\be
\ba{lll}
\partial_{\mu i} = 0 \,,\quad&\quad\partial_{ij} = 0 \,,\quad&\quad\partial_{\mu\nu}\neq 0\,,
\ea
\label{f3section}
\ee
where $\mu, \nu = 1, 2, 3$ and $i, j = 4, 5,\cdots, N$.  In this case, we may dualize the  nontrivial  three coordinates, using  a three-dimensional Levi-Civita symbol, 
$\varepsilon_{123}\equiv1$,
\be
\ba{ll}
\tilde{x}_{\mu} \equiv \half\varepsilon_{\mu\nu\rho} x^{\nu\rho} \,, \quad&\quad
\tilde{\partial}^{\mu}\tilde{x}_{\nu}=\delta^{\mu}_{~\nu}\,. 
\ea
\ee
\end{enumerate}
For a triplet of arbitrary functions,  we note~\cite{Blair:2013gqa}
\be
\ba{ll}
\partial_{[ab} \Phi\partial_{c][d} \Phi^{\prime}\partial_{ef]}\Phi^{\prime\prime}= 0~~~\mbox{on}~~~\Sigma_{N-1}\,,\quad&\quad
\partial_{[ab} \Phi\partial_{c][d} \Phi^{\prime}\partial_{ef]}\Phi^{\prime\prime}\neq 0~~~\mbox{on}~~~\Sigma_{3}\,.
\ea
\ee
Since this is an  $\SLN$ covariant statement, the two sections, $\Sigma_{N-1}$ and $\Sigma_{3}$, are  duality  inequivalent.  More than one solution to a section condition    has been  also reported  in  EFT~\cite{Hohm:2013vpa,Hohm:2013uia}.

\item \textit{{\textbf{Riemannian reductions.}}} 

\begin{enumerate}
\item To perform the Riemannian reduction  to  $\Sigma_{N-1}$~(\ref{fMsection}), we parametrize the U-metric in terms of $(N-1)$-dimensional   Riemannian metric, $g_{\alpha\beta}$,  a vector, $v^{\alpha}$,  and a scalar, $\phi$~\cite{Park:2013gaj},
\be
\ba{ll}
 M_{ab} = 
 \left( \begin{array}{cc}
 \frac{g_{\alpha\beta}}{\sqrt{|g|}} & v_{\alpha} \\
		  v_{\beta} & \sqrt{|g|} \left(- e^{\phi} + v^{2}\right)
                 \end{array} \right) \,,\quad&\quad 
               |M|^{\frac{1}{4-N}}=e^{\frac{1}{4-N}\phi}\sqrt{|g|}\,.
\ea
\label{fPARAN}
\ee
The U-gravity scalar curvature~(\ref{fRicciScalar}) reduces upon the section, $\Sigma_{N-1}$, to
\be
\left.S\right|_{\Sigma_{N-1}}=2e^{-\phi}\left[ R_{g}-\textstyle{\frac{(N-3)(3N-8)}{4(N-4)^{2}}}\partial_{\alpha}\phi\partial^{\alpha}\phi+\textstyle{\frac{N-2}{N-4}}\Delta\phi+\half e^{-\phi}\left(\trd_{\alpha}v^{\alpha}\right)^{2}\right]\,.
\label{fREDUCEDN}
\ee
The vector field can be dualized to an $(N-2)$-form  potential.

\item For the Riemannian reduction to  $\Sigma_{3}$~(\ref{f3section}),  we parametrize (the inverse of) the U-metric, employing     `dual' upside-down  notations~\cite{Blair:2013gqa},  
\be
\ba{ll}
M^{ab}=\left(\ba{cc}
\frac{\tg^{\mu\nu}}{\sqrt{|\tg|}}~&~
-\tv^{j\mu}
\\
-\tv^{i\nu}~&~\sqrt{|\tg|}(e^{-\tphi}\tilde{\cM}^{ij}+\tv^{i\lambda}\tv^{j}{}_{\lambda})
\ea
\right)\,,\quad&\quad 
|M|^{\frac{1}{4-N}}=e^{\frac{N-3}{4-N}\tphi}\sqrt{|\tg|}\,.
\ea
\label{fPARA3}
\ee
The U-gravity scalar curvature~(\ref{fRicciScalar}) reduces upon the section, $\Sigma_{3}$, to
\be
 \left.S\right|_{\Sigma_{3}}= -2R_{\tg} + \textstyle{\frac{(N-3)(3N-8)}{2(N-4)^{2}}} 
 \tpartial^{\mu}\tphi \tpartial_{\mu}\tphi-\textstyle{\frac{4(N-3)}{N-4}}\tilde{\Delta}\tphi
 -\half\tpartial^{\mu}\tilde{\cM}_{ij} \tpartial_{\mu}\tilde{\cM}{}^{ij}
+ e^{\tphi} \tilde{\cM}_{ij}\tilde{\trd}{}^{\mu}\tv^{i}{}_{\mu}\tilde{\trd}{}^{\nu}\tv^{j}{}_{\nu}\,,
\label{fREDUCED3}
\ee
which manifests  $\,\mathbf{SL}(N{-3})$ \underline{S-duality}.
\end{enumerate}

\item \textit{{\textbf{Non-Riemannian backgrounds.}}}    When the upper left $(N-1)\times(N-1)$ block of the U-metric is degenerate --where $\frac{g_{\alpha\beta}}{\sqrt{|g|}}$ is positioned in (\ref{fPARAN})--  the Riemannian metric ceases to exist  upon  $\Sigma_{N-1}$. Nevertheless,    $\SLN$ U-gravity has no problem with describing   such a non-Riemannian background, as long as  the whole $N\times N$  U-metric  is non-degenerate.  Similarly upon $\Sigma_{3}$,  U-gravity may allow the upper left $3\times 3$ block of 
the inverse of the  U-metric~(\ref{fPARA3}) to be degenerate (See section~\ref{SECREDUCTION} for further discussion with  examples).\footnote{Consult  also \cite{Lee:2013hma}  for  a parallel  discussion in DFT.}

\end{itemize}
~\\

\section{Exposition\label{SECFORM}} 
In this section we provide     detailed  exposition  of  the   main results   listed   in   section~\ref{SECSUMMARY}.     All the mathematical analyses  are parallel to those in the DFT-geometry of  \cite{Jeon:2010rw,Jeon:2011cn,Park:2012xn,Park:2013mpa,Lee:2013hma}.

\subsection{Equivalence between the  coordinate gauge symmetry and the  section condition \label{SUBSECGauge}}
Here, following a parallel argument  in DFT~\cite{Lee:2013hma},  we show  the equivalence between  the  coordinate gauge symmetry invariance~(\ref{fTensorCGS}), 
\be
\ba{ll}
\Phi(x+s\Delta)=\Phi(x)\,,\quad&\quad\Delta^{ab}=\textstyle{\frac{1}{(N-4)!}}\epsilon^{abc_{1}\cdots c_{N-4}de}\phi_{c_{1}\cdots c_{N-4}}\partial_{de}\varphi\,,
\ea
\label{TensorCGS}
\ee
and the section condition~(\ref{fseccon2}),
\begin{eqnarray}
&\partial_{[ab}\Phi\partial_{cd]}\Phi^{\prime}=
\half\partial_{[ab}\Phi\partial_{c]d}\Phi^{\prime}-\half\partial_{d[a}\Phi\partial_{bc]}\Phi^{\prime}\seceq 0&\quad(\mbox{strong~~constraint})\,,\label{seccon2}\\
&\partial_{[ab}\partial_{cd]}\Phi=\partial_{[ab}\partial_{c]d}\Phi\seceq 0&\quad(\mbox{weak~~constraint})\,.\label{seccon1}
\end{eqnarray} 
Note that, in (\ref{TensorCGS}) we put a continuous real parameter,  $s$, in order to control the shift.

\noindent First of all, from   the standard  series expansion of $\Phi(x+s\Delta)$ in $s$,   it is clear that the strong  constraint, (\ref{seccon2}),  implies the invariance~(\ref{TensorCGS}). The converse is also true: taking derivative at $s=0$, we get
\be
0=\left.\frac{\rmd~}{\rmd s}\Phi(x+s\Delta)\right|_{s=0}=\half\Delta^{ab}\partial_{ab}\Phi=\textstyle{\frac{1}{2(N-4)!}}\epsilon^{c_{1}\cdots c_{N-4}deab}\phi_{c_{1}\cdots c_{N-4}}\partial_{de}\varphi\partial_{ab}\Phi\,.
\ee
This shows that  the invariance~(\ref{TensorCGS}) indeed implies  the strong constraint~(\ref{seccon2}).  Further, from  the strong constraint, it follows that  the following  $\frac{N(N-1)}{2}\times\frac{N(N-1)}{2}$  matrix is nilpotent, 
\be
\ba{ll}
\cK^{ab}{}_{cd}=\textstyle{\frac{1}{(N-4)!}}\epsilon^{abe_{1}\cdots e_{N-4}fg}\phi_{e_{1}\cdots e_{N-4}}\partial_{fg}\partial_{cd}\varphi\,,\quad&\quad \half \cK^{ab}{}_{cd}\cK^{cd}{}_{ef}=0\,.
\ea
\ee
Since any  nilpotent matrix is traceless\footnote{All the diagonal elements of  the Jordan normal form of a nilpotent matrix are trivial~\cite{Lee:2013hma}.}, we have
\be
\cK^{ab}{}_{ab}=\textstyle{\frac{1}{(N-4)!}}\epsilon^{e_{1}\cdots e_{N-4}abfg}\phi_{e_{1}\cdots e_{N-4}}\partial_{ab}\partial_{fg}\varphi=0\,,
\ee
which leads to the weak constraint~(\ref{seccon1}), 
\be
\partial_{[ab}\partial_{cd]}\varphi\seceq 0\,.\label{seccon1p}
\ee
In this way,  the strong constraint~(\ref{seccon2})  implies the weak constraint~(\ref{seccon1}), and is actually  equivalent to the coordinate gauge symmetry invariance~(\ref{TensorCGS}).   This completes our proof.

\subsection{Projection operator}
The eight-index projection operator~(\ref{fprojection}),
\be
\ba{ll}
\cP_{abcd}{}^{klmn}=&
\half\delta_{[a}^{~[k}\delta_{b]}^{~l]}\delta_{[c}^{~[m}\delta_{d]}^{~n]}
+\half\delta_{[c}^{~[k}\delta_{d]}^{~l]}\delta_{[a}^{~[m}\delta_{b]}^{~n]}
+\half M_{c[a}\delta_{b]}^{~m}M^{n[k}\delta_{d}^{~l]}
-\half M_{c[a}\delta_{b]}^{~[k}M^{l]n}\delta_{d}^{~m}\\
{}&~+\textstyle{\frac{1}{N-2}}\left(
\delta_{[a}^{~n}M_{b][c}M^{m[k}\delta_{d]}^{~l]}
+\delta_{[c}^{~n}M_{d][a}M^{m[k}\delta_{b]}^{~l]}
-M_{c[a}M_{b]d}M^{m[k}M^{l]n}\right)\,,
\ea
\label{projection}
\ee
satisfies the `projection' property,
\be
\cP_{abcd}{}^{pqrs}\cP_{pqrs}{}^{klmn}=\cP_{abcd}{}^{klmn}\,.
\label{projection2}
\ee
The verification of this identity    requires straightforward yet tedious computations,  which can  be simplified by noting  symmetric properties, 
\be
\ba{ll}
\cP_{abcd}{}^{klmn}=\cP_{[ab]cd}{}^{[kl]mn}\,,\quad&\quad
\cP_{ab[cd]}{}^{klmn}=\cP_{cd[ab]}{}^{klmn}\,,
\ea
\label{projectionsym}
\ee
and   `trace' properties, 
\be
\ba{l}
\cP^{s}{}_{asb}{}^{klmn}=\half(N-2)\delta_{(a}^{~m}M^{n[k}\delta_{b)}^{~l]}
-\textstyle{\frac{N}{2(N-2)}}M_{ab}M^{m[k}M^{l]n}\,,\\
\cP^{s}{}_{abs}{}^{klmn}=\delta_{(a}^{~m}M^{n[k}\delta_{b)}^{~l]}
+\textstyle{\frac{N-1}{N-2}}M_{ab}M^{m[k}M^{l]n}\,,\\
\cP^{rs}{}_{rs}{}^{klmn}=-\left(\textstyle{\frac{N^{2}-2N+2}{N-2}}\right)M^{m[k}M^{l]n}\,,\\
\cP_{abs}{}^{sklmn}=0\,.
\ea
\label{traceP}
\ee
The traces are related to each other by
\be
\cP^{s}{}_{asb}{}^{klmn}=\half(N-2)\cP^{s}{}_{abs}{}^{klmn}
+\half M_{ab}\cP^{rs}{}_{rs}{}^{klmn}\,.
\ee
It is also useful to note
\be
\cP_{[abc]d}{}^{klmn}=\cP_{[abcd]}{}^{[klmn]}=
\delta_{[a}^{~[k}\delta_{b}^{~l}\delta_{c}^{~m}\delta_{d]}^{~n]}\,.
\label{3aP}
\ee
As we shall see below, the projection operator  plays  crucial roles in U-gravity.\footnote{The construction  of the projection operator~(\ref{projection})  is one of the major   improvements made in this paper  compared to the  previous  work on $\SLf$ U-geometry~\cite{Park:2013gaj}. An  operator therein, called $J_{abcd}{}^{klmn}$,   is consistently related to the projection operator by
\[
\ba{ll}
J_{abcd}{}^{klmn}&=
\half\delta_{[a}^{~[k}\delta_{b]}^{~l]}\delta_{[c}^{~[m}\delta_{d]}^{~n]}
+\half\delta_{[c}^{~[k}\delta_{d]}^{~l]}\delta_{[a}^{~[m}\delta_{b]}^{~n]}
+\textstyle{\frac{1}{N-2}}\left(
\delta_{[a}^{~n}M_{b][c}M^{m[k}\delta_{d]}^{~l]}
+\delta_{[c}^{~n}M_{d][a}M^{m[k}\delta_{b]}^{~l]}\right)\\
{}&=\cP_{abcd}{}^{klmn}-M_{c[a}\cP^{s}{}_{b]ds}{}^{klmn}-\textstyle{\frac{N}{N^{2}-2N+2}}M_{c[a}M_{b]d}\cP^{rs}{}_{rs}{}^{klmn}\,.
\ea
\]}  Compared to the ordinary Riemannian geometry, the existence of a projection operator and its  key   role appear   to be  novel distinct  features    of  the   extended-yet-gauged spacetime geometries,  such as  DFT-geometry in \cite{Jeon:2010rw,Jeon:2011kp,Jeon:2011cn,Jeon:2011vx,Jeon:2011sq,Jeon:2012kd,Jeon:2012hp} and the present $\SLN$ U-gravity.

\subsection{Compatibility of the semi-covariant derivative}
Here we discuss the compatibilities of the semi-covariant derivative,  firstly with the generalized Lie derivative, secondly with the generalized bracket, and lastly with the U-metric.

\noindent Specifically, we start by postulating   the generalized Lie derivative and the semi-covariant derivative to take  the following forms,
\be
\ba{ll}
\hcL_{X}T^{a_{1}a_{2}\cdots a_{p}}{}_{b_{1}b_{2}\cdots b_{q}}:=\!\!&
\half X^{cd}\partial_{cd}T^{a_{1}a_{2}\cdots a_{p}}{}_{b_{1}b_{2}\cdots b_{q}}
+\alpha(p,q,\omega)\partial_{cd}X^{cd}T^{a_{1}a_{2}\cdots a_{p}}{}_{b_{1}b_{2}\cdots b_{q}}\\
{}&-\sum_{i=1}^{p}T^{a_{1}\cdots c\cdots a_{p}}{}_{b_{1}b_{2}\cdots b_{q}}\partial_{cd}X^{a_{i}d}
+\sum_{j=1}^{q}\partial_{b_{j}d}X^{cd}T^{a_{1}a_{2}\cdots a_{p}}{}_{b_{1}\cdots c\cdots b_{q}}\,,\\
\na_{cd}T^{a_{1}a_{2}\cdots a_{p}}{}_{b_{1}b_{2}\cdots b_{q}}:=\!\!&
\partial_{cd}T^{a_{1}a_{2}\cdots a_{p}}{}_{b_{1}b_{2}\cdots b_{q}}
+\bar{\alpha}(p,q,\omega)\Gamma_{cde}{}^{e}T^{a_{1}a_{2}\cdots a_{p}}{}_{b_{1}b_{2}\cdots b_{q}}\\
{}&~-\sum_{i=1}^{p}T^{a_{1}\cdots e\cdots a_{p}}{}_{b_{1}b_{2}\cdots b_{q}}\Gamma_{cde}{}^{a_{i}}+\sum_{j=1}^{q}\Gamma_{cdb_{j}}{}^{e}T^{a_{1}a_{2}\cdots a_{p}}{}_{b_{1}\cdots e\cdots b_{q}}\,.
\ea
\label{gLsemicov}
\ee
Here, $\alpha(p,q,\omega)$ and  $\bar{\alpha}(p,q,\omega)$ are yet-undetermined total weights   which may depend on $p,q,\omega$, \textit{i.e.} the numbers of upper, lower indices and the  extra weight.   Below, in section~\ref{SECcomp1}, by demanding the compatibility   with the generalized Lie derivative, we shall fix the dependency  and derive the final expression, 
\be
\alpha(p,q,\omega)=\bar{\alpha}(p,q,\omega)=\half(\half p- \half q+\omega)\,,
\label{alphabar}
\ee 
which is linear in $p,q,\omega$ and remarkably  independent of $N$. This result will, in particular,  ensure that  both  the generalized Lie derivative and    the semi-covariant  derivative  annihilate the   Kronecker delta symbol,
\be
\ba{ll}
\hcL_{X}\delta^{a}_{~b}=0\,,\quad&\quad\na_{cd}\delta^{a}_{~b}=0\,.
\ea
\ee
Further, while the extra weight of the U-metric is trivial, its determinant, $M\equiv\det(M_{ab})$,  acquires  an extra weight, $\omega=4-N$, since under diffeomorphism, it transforms as
\be
\delta_{X}M=\half X^{ab}\partial_{ab}M+\half(4-N)\partial_{ab}X^{ab}M\,.
\label{gLNdet}
\ee
This implies    that  \textit{the duality invariant integral measure}  with unit extra weight is   
\be
|M|^{\frac{1}{4-N}}\,.
\label{measurep}
\ee
~\\
\noindent It is instructive to note that, irrespective of the choice of $\alpha(p,q,\omega)$, upon the section condition, the commutator of the  generalized Lie derivative is closed by a generalized bracket~\cite{Berman:2011cg},
\begin{eqnarray}
&&{}\left[\hcL_{X},\hcL_{Y}\right]T^{a_{1}a_{2}\cdots a_{p}}{}_{b_{1}b_{2}\cdots b_{q}}=\hcL_{[X,Y]_{\rmG}}T^{a_{1}a_{2}\cdots a_{p}}{}_{b_{1}b_{2}\cdots b_{q}}\,,\label{comgL}\\
&&{}[X,Y]^{ab}_{\rmG}=\half X^{cd}\partial_{cd}Y^{ab}-\textstyle{\frac{3}{2}}X^{[ab}\partial_{cd}Y^{cd]}-\half Y^{cd}\partial_{cd}X^{ab}+\textstyle{\frac{3}{2}}Y^{[ab}\partial_{cd}X^{cd]}\,.
\label{gB}
\end{eqnarray}
Further, it is obvious  from this expression that the generalized bracket satisfies up to the section condition,
\be
\ba{l}
[X,Y]^{ab}_{\rmG}\partial_{ab}\Phi=\half (X^{cd}\partial_{cd}Y^{ab}-
Y^{cd}\partial_{cd}X^{ab})\partial_{ab}\Phi\,,\\
\partial_{ab}\left([X,Y]^{ab}_{\rmG}\right)=\half\left(X^{cd}\partial_{cd}\partial_{ab}Y^{ab}-Y^{cd}\partial_{cd}\partial_{ab}X^{ab}\right)\,.
\ea
\ee

\subsubsection{Compatibility with the generalized Lie derivative\label{SECcomp1}}
If we replace all the ordinary derivatives by semi-covariant derivatives in the definition of the generalized Lie derivative expressed  in (\ref{gLsemicov}), we get
\be
\ba{ll}
\left[\hcL_{X}(\na)-\hcL_{X}(\partial)\right]T^{a_{1}\cdots a_{p}}{}_{b_{1}\cdots b_{q}}=&\!\!\!X^{cd}\left[(\half\bar{\alpha}+\alpha\bar{\beta})\Gamma_{cde}{}^{e}+2\alpha\Gamma_{e[cd]}{}^{e}\right]T^{a_{1}\cdots a_{p}}{}_{b_{1}\cdots b_{q}}\\
{}&\!\!\!\!\!-\sum_{i=1}^{p}T^{a_{1}\cdots e\cdots a_{p}}{}_{b_{1}\cdots b_{q}}\left[\textstyle{\frac{3}{2}}X^{cd}\Gamma_{[cde]}{}^{a_{i}}+X^{a_{i}d}(\bar{\beta}\Gamma_{edc}{}^{c}-\Gamma_{ecd}{}^{c})\right]\\
{}&\!\!\!\!\!+\sum_{j=1}^{q}\left[\textstyle{\frac{3}{2}}X^{cd}\Gamma_{[cdb_{j}]}{}^{e}+X^{ed}(\bar{\beta}\Gamma_{b_{j}dc}{}^{c}-\Gamma_{b_{j}cd}{}^{c})\right]
T^{a_{1}\cdots a_{p}}{}_{b_{1}\cdots e\cdots b_{q}}\,,
\ea
\label{comp1}
\ee
where we set for the parameter, $X^{ab}$,
\be
\bar{\beta}\equiv\bar{\alpha}(2,0,0)\,.
\label{barbeta}
\ee
The compatibility of the semi-covariant derivative with the generalized Lie derivative means that the right hand side of (\ref{comp1}) should vanish algebraically.  In order to achieve this, it is  required  that the four-index quantity, $\Gamma_{[abc]}{}^{d}$, should be, at least,   related  to the   two-index quantities, $\Gamma_{eab}{}^{e}$ and $\Gamma_{abe}{}^{e}$. There is one unique such an ansatz which is  self-consistent,\footnote{The division by ${N-2}$ in (\ref{ansatz}) needs not cause any alarm to exclude the case of ${N=2}$, since after all we shall have $\Gamma_{[abc]}{}^{d}=0$~(\ref{fGabc}).}
\be
\ba{ll}
\Gamma_{[abc]}{}^{d}=\textstyle{\frac{1}{N-2}}
\hat{\Gamma}_{[ab}\delta^{~d}_{c]}\,,\quad&\quad
\hat{\Gamma}_{ab}=3\Gamma_{[abe]}{}^{e}\,.
\ea
\label{ansatz}
\ee
Note that the  left and right hand sides of this ansatz  share the same anti-symmetric properties, and also that the contractions of   the two indices, one lower and the other  upper (for example $c$ and $d$\,), agree.\\
Assuming the ansatz~(\ref{ansatz}), the expression~(\ref{comp1}) reduces to
\be
\ba{l}
\left[\hcL_{X}(\na)-\hcL_{X}(\partial)\right]T^{a_{1}\cdots a_{p}}{}_{b_{1}\cdots b_{q}}\\
=X^{cd}\left[(\half\bar{\alpha}+\alpha\bar{\beta}
+\textstyle{\frac{q-p}{2(N-2)}})\Gamma_{cde}{}^{e}+
(2\alpha+\textstyle{\frac{q-p}{N-2}})\Gamma_{e[cd]}{}^{e}\right]T^{a_{1}\cdots a_{p}}{}_{b_{1}\cdots b_{q}}\\
\quad-\sum_{i=1}^{p}
X^{a_{i}d}
\left[(\bar{\beta}-\textstyle{\frac{1}{N-2}})\Gamma_{edc}{}^{c}
+\textstyle{\frac{N-4}{N-2}}\Gamma_{c[ed]}{}^{c}
+\Gamma_{c(ed)}{}^{c}\right]
T^{a_{1}\cdots e\cdots a_{p}}{}_{b_{1}\cdots b_{q}}\\
\quad+\sum_{j=1}^{q}X^{ed}
\left[(\bar{\beta}-\textstyle{\frac{1}{N-2}})\Gamma_{b_{j}dc}{}^{c}
+\textstyle{\frac{N-4}{N-2}}\Gamma_{c[b_{j}d]}{}^{c}
+\Gamma_{c(b_{j}d)}{}^{c}
\right]T^{a_{1}\cdots a_{p}}{}_{b_{1}\cdots e\cdots b_{q}}\,.
\ea
\label{comp1p}
\ee
Now, each line above should vanish separately. More precisely,  with the skew-symmetry,  $\Gamma_{abc}{}^{d}=\Gamma_{[ab]c}{}^{d}$, we should require
\begin{eqnarray}
&&(\half\bar{\alpha}+\alpha\bar{\beta}
+\textstyle{\frac{q-p}{2(N-2)}})\Gamma_{abc}{}^{c}+
(2\alpha+\textstyle{\frac{q-p}{N-2}})\Gamma_{c[ab]}{}^{c}=0\,,
\label{con1}\\
&&(\bar{\beta}-\textstyle{\frac{1}{N-2}})\Gamma_{abc}{}^{c}
+\textstyle{\frac{N-4}{N-2}}\Gamma_{c[ab]}{}^{c}=0\,,
\label{con2}\\
&&\Gamma_{c(ab)}{}^{c}=0\,.\label{con3}
\end{eqnarray}
Equation (\ref{con2}) gives an expression, $\Gamma_{c[ab]}{}^{c}=\frac{1-(N-2)\bar{\beta}}{N-4}\Gamma_{abc}{}^{c}$. Substituting this into (\ref{con1}), we get
\be
\left[\half\bar{\alpha}+\alpha\bar{\beta}
+\textstyle{\frac{q-p}{2(N-2)}}+
(2\alpha+\textstyle{\frac{q-p}{N-2}})\textstyle{\frac{1-(N-2)\bar{\beta}}{N-4}}\right]\Gamma_{abc}{}^{c}=0\,.
\label{con4}
\ee
There is a good reason for the contraction, $\Gamma_{abc}{}^{c}$,  to be nontrivial: as we shall discuss   more  in section~\ref{SECcomp3},  the compatibility  of the semi-covariant derivative  with  the U-metric, and hence  with  its determinant,   implies  for some value\footnote{In fact, $\omega^{\ast}=4-N$ from (\ref{alphabar}) and  (\ref{gLNdet}).} of $\omega^{\ast}$,
\be
\na_{ab}M=\partial_{ab}M+\bar{\alpha}(0,0,\omega^{\ast})\Gamma_{abc}{}^{c}M\equiv0\,.
\label{compdet}
\ee  
For this to hold, $\Gamma_{abc}{}^{c}$ should not vanish in general.   Thus, Eq.(\ref{con4}) tells us
\be
\half\bar{\alpha}+\alpha\bar{\beta}
+\textstyle{\frac{q-p}{2(N-2)}}+
(2\alpha+\textstyle{\frac{q-p}{N-2}})\textstyle{\frac{1-(N-2)\bar{\beta}}{N-4}}=0\,.
\label{con5}
\ee
In particular,  for  the special case of $p=2$, $q=0$, $\omega=0$,  this reduces to
\be
(N\bar{\beta}-2)(2\beta-1)=0\,,
\ee
where, like (\ref{barbeta}), we set ${\beta}\equiv{\alpha}(2,0,0)$. Hence, we have  either $\bar{\beta}=\frac{2}{N}$ or $\beta=\half$. If $\bar{\beta}=\frac{2}{N}$,  Eq.(\ref{con5}) would get simplified to  give  $\bar{\alpha}(p,q,\omega)=\frac{p-q}{N}$. However, this is not a desired result. In order to meet the  compatibility   with the U-metric determinant~(\ref{compdet}), $\bar{\alpha}(p,q,\omega)$  must depend nontrivially on $\omega$ rather than being independent of it.   Therefore, we should choose $\beta=\half$. 

\noindent Now,  rather than trying  to look for the most general solution,  for simplicity  we  focus on the case of  $\beta=\bar{\beta}=\half$ and search for a `linear' solution. Then,  Eq.(\ref{con5})  implies a more generic  equality,  $\alpha=\bar{\alpha}$, and  naturally we are   lead to   the  final  expression for the total weight, \textit{i.e.~}(\ref{alphabar}). Further, (\ref{con1}) and (\ref{con2}) reduce to
\be
\Gamma_{abc}{}^{c}+2\Gamma_{c[ab]}{}^{c}=\hat{\Gamma}_{ab}=0\,,
\label{hatGv}
\ee
and thus, from (\ref{ansatz}) and (\ref{con3}), we arrive at the conclusion: the  conditions    for the  compatibility  of the semi-covariant derivative with the generalized Lie derivative are
\be
\ba{lll}
\Gamma_{[abc]}{}^{d}=0\,,\quad&\quad\Gamma_{c(ab)}{}^{c}=0\,,\quad&\quad
\alpha(p,q,\omega)=\bar{\alpha}(p,q,\omega)=\half(\half p- \half q+\omega)\,.
\label{finalp}
\ea
\ee

\subsubsection{Compatibility with the generalized bracket\label{SECcomp2}}
If we replace all the ordinary derivatives by semi-covariant derivatives in the definition of the generalized bracket~(\ref{gB}), we get, in a similar fashion to (\ref{comp1}),
\be
\ba{ll}
[X,Y]^{ab}_{\rmG}(\na)-[X,Y]^{ab}_{\rmG}(\partial)=&
\half(Y^{ab}X^{cd}-X^{ab}Y^{cd})(\bar{\beta}\Gamma_{cde}{}^{e}+\Gamma_{e[cd]}{}^{e})\\
{}&+(X^{ac}Y^{bd}-Y^{ac}X^{bd})\Gamma_{e(cd)}{}^{e}\\
{}&+\textstyle{\frac{3}{2}}\Gamma_{[cde]}{}^{[a}Y^{b]e}X^{cd}-\textstyle{\frac{3}{2}}\Gamma_{[cde]}{}^{[a}X^{b]e}Y^{cd}\,,
\ea
\ee
which further reduces, with the ansatz (\ref{ansatz}),  to
\be
\ba{ll}
[X,Y]^{ab}_{\rmG}(\na)-[X,Y]^{ab}_{\rmG}(\partial)=&
\half(Y^{ab}X^{cd}-X^{ab}Y^{cd})\left[(\bar{\beta}-\textstyle{\frac{1}{N-2}})\Gamma_{cde}{}^{e}+\textstyle{\frac{N-4}{N-2}}\Gamma_{e[cd]}{}^{e})\right]\\
{}&+(X^{ac}Y^{bd}-Y^{ac}X^{bd})\Gamma_{e(cd)}{}^{e}\,.
\ea
\ee
In order to meet the compatibility, each line should vanish separately. Hence, we require 
\be
\ba{ll}
(\bar{\beta}-\textstyle{\frac{1}{N-2}})\Gamma_{abc}{}^{c}
+\textstyle{\frac{N-4}{N-2}}\Gamma_{c[ab]}{}^{c}=0\,,\quad&\quad
\Gamma_{c(ab)}{}^{c}=0\,,
\ea
\ee
which in fact coincide with (\ref{con2}) and (\ref{con3}).  Thus, putting $\bar{\beta}\equiv\frac{1}{2}$, we  re-derive  (\ref{hatGv}) and, from (\ref{ansatz}), we arrive at the same conditions  as  before  for the connection (\ref{finalp}), 
\be
\ba{ll}
\Gamma_{[abc]}{}^{d}=0\,,\quad&\quad\Gamma_{c(ab)}{}^{c}=0\,.
\label{finalpp}
\ea
\ee

\subsubsection{Compatibility with the U-metric\label{SECcomp3}}
Having fixed the total weight to be $\half(\half p- \half q+\omega)$ as (\ref{alphabar}), 
the compatibility  of the semi-covariant derivative  with the U-metric  reads
\be
\na_{ab}M_{cd}=\partial_{ab}M_{cd}-\half\Gamma_{abe}{}^{e}M_{cd}+2\Gamma_{ab(cd)}=0\,.
\label{Metriccomp}
\ee
Contracting   $c$ and $d$ indices we get
\be
\Gamma_{abe}{}^{e}=\textstyle{\frac{2}{N-4}}\partial_{ab}\ln\left|M\right|\,.
\label{Gammaee}
\ee
Thus, the metric compatibility~(\ref{Metriccomp}) is equivalent to
\be
A_{abcd}:=\Gamma_{ab(cd)}=-\half\partial_{ab}M_{cd}+\textstyle{\frac{1}{2(N-4)}}M_{cd}\partial_{ab}\ln\left|M\right|\,.
\label{defA}
\ee
It is useful to note
\be
\ba{l}A_{abe}{}^{e}=\Gamma_{abe}{}^{e}=\textstyle{\frac{2}{N-4}}\partial_{ab}\ln|M|=\textstyle{\frac{2}{N-4}}M^{ef}\partial_{ab}M_{ef}\,,\\
\partial_{ab}M_{cd}=-2A_{abcd}+\half A_{abe}{}^{e}M_{cd}\,,\\ 
\partial_{ab}M^{cd}=2A_{ab}{}^{cd}-\half A_{abe}{}^{e}M^{cd}\,.
\ea
\label{Auseful}
\ee

\subsection{Determining the  connection uniquely}
Here, we derive the  connection (\ref{fGamma}), 
\be
\ba{ll}
\Gamma_{abcd}&=A_{abcd}+\half(A_{acbd}-A_{adbc}+A_{bdac}-A_{bcad})\\
{}&\quad+\textstyle{\frac{1}{N-2}}\left(M_{ac}A^{e}{}_{(bd)e}-M_{ad}A^{e}{}_{(bc)e}+M_{bd}A^{e}{}_{(ac)e}-M_{bc}A^{e}{}_{(ad)e}\right)\,,
\ea
\label{connectionGamma}
\ee
as the unique solution to the five constraints, (\ref{fcompM}), (\ref{fGab}), (\ref{fGabc}), (\ref{fGcc}) and (\ref{flastcon}). The connection can be  rewritten,
\be
\Gamma_{abcd}=B_{[ab]cd}+\half(B_{acbd}-B_{adbc}+B_{bdac}-B_{bcad})\,,
\ee
if we set
\be
B_{abcd}:=A_{abcd}+\textstyle{\frac{2}{N-2}} M_{ab}A^{e}{}_{(cd)e}\,.
\ee
~~\\
\noindent We start by  recalling   the five conditions for  the  connection,
\begin{eqnarray}
&&\Gamma_{ab(cd)}=A_{abcd}\,,\label{compM}\\
&&\Gamma_{(ab)c}{}^{d}=0\,,\label{Gab}\\
&&\Gamma_{[abc]}{}^{d}=0\,,\label{Gabc}\\
&&\Gamma_{c(ab)}{}^{c}=0\,,\label{Gcc}\\
&&\cP_{abcd}{}^{efgh}\Gamma_{efgh}=0\,.\label{lastcon}
\end{eqnarray}
The first condition~(\ref{compM}) is equivalent to the metric compatibility, $\na_{ab}M_{cd}=0$, as discussed in section~\ref{SECcomp3}.  The second condition~(\ref{Gab}) is natural, from the skew-symmetric property of the coordinates, $x^{(ab)}=0$ and hence  $\partial_{(ab)}=\na_{(ab)}=0$. The next two relations,  (\ref{Gabc}) and (\ref{Gcc}), ensure  the compatibilities  with   the generalized Lie derivative and also with the generalized bracket, as discussed  in  sections, \ref{SECcomp1} and \ref{SECcomp2}.   The last condition~(\ref{lastcon}) is a projection  property  which we deliberately impose  in order to fix the  connection uniquely.  We may view the three constraints,   (\ref{Gabc}), (\ref{Gcc}) and (\ref{lastcon}), as the `torsionless' conditions. These are all  --including the projection condition--     analogous  to the DFT-geometry of \cite{Jeon:2011cn}.\\

\noindent While the first condition, (\ref{compM}), fixes the symmetric part of the connection, the remaining ones should  determine the anti-symmetric part,
\be
\cX_{abcd}=\cX_{[ab][cd]}:=\Gamma_{ab[cd]}\,,
\ee
satisfying
\be
\Gamma_{abcd}=A_{abcd}+\cX_{abcd}\,.
\ee
First of all, it follows  from (\ref{compM}), (\ref{Gcc}),
\be
\Gamma^{c}{}_{acb}+\Gamma^{c}{}_{bca}=4A^{c}{}_{(ab)c}\,,
\label{Gcc2}
\ee
and also from (\ref{Gab}), (\ref{Gcc}),
\be
\ba{ll}
\Gamma^{cd}{}_{dc}=0\,,\quad&\quad\Gamma^{cd}{}_{cd}=0\,.
\ea
\label{Gcdcd}
\ee
Further,  from \textit{e.g.~}$\Gamma_{[abc]d}-\Gamma_{[abd]c}+\Gamma_{[bcd]a}-\Gamma_{[acd]b}=0$, we get
\be
\cX_{abcd}-\cX_{cdab}=2A_{a[cd]b}-2A_{b[cd]a}\,.
\ee
We then only need to determine
\be
\cY_{abcd}:=\half\left(\cX_{abcd}+\cX_{cdab}\right)=\half\left(\Gamma_{[ab][cd]}+\Gamma_{[cd][ab]}\right)\,,
\ee
which satisfies, by construction,   symmetric properties,
\be
\ba{ll}
\cY_{abcd}=\cY_{[ab][cd]}\,,\quad&\quad \cY_{abcd}=\cY_{cdab}\,,
\ea
\label{Ysym}
\ee
and contributes to the connection through
\be
\Gamma_{abcd}=A_{abcd}+\half(A_{acdb}-A_{adcb}-A_{bcda}+A_{bdca})+\cY_{abcd}\,.
\ee
Now, all the constraints except the last one~(\ref{lastcon}),  boil down to
\be
\ba{ll}
\cY_{[abc]d}=0\,,\quad&\quad \cY^{c}{}_{acb}=A^{c}{}_{(ab)c}\,.
\ea
\label{Yrel}
\ee
On the other hand, the last projection condition  (\ref{lastcon}) fixes $Y_{abcd}$ uniquely, 
\be
\cY_{abcd}=\textstyle{\frac{1}{N-2}}\left(M_{ac}A^{e}{}_{(bd)e}-M_{bc}A^{e}{}_{(ad)e}+M_{bd}A^{e}{}_{(ac)e}-M_{ad}A^{e}{}_{(bc)e}\right)\,.
\label{Ysol}
\ee
It is straightforward to check for consistency that, $\cY_{abcd}$ given in (\ref{Ysol}) indeed  satisfies the relations (\ref{Yrel}) and also  the (anti-)symmetric properties (\ref{Ysym}). Alternatively,  one may well guess the expression (\ref{Ysol}) as \textit{a} solution of (\ref{Ysym}) and (\ref{Yrel}), \textit{i.e.} \textit{a} solution that  can be readily   constructed in terms of the symmetric two-index objects,  $M_{ab}$ and $A^{e}{}_{(ab)e}$.   The last condition (\ref{lastcon}) then ensures it to be \textit{the} only  solution.\\

\noindent  Following the  method in \cite{Park:2013gaj},  the uniqueness can be also verified directly.    First, it is straightforward to check that  the connection given in (\ref{connectionGamma})  satisfies all the five conditions, (\ref{compM}) --- (\ref{lastcon}).  On the other hand, if the most general solution of them might contain  an extra piece, say $\Upsilon_{abcd}$, the first four conditions, (\ref{compM}) --- (\ref{Gcc}), imply 
\be
\ba{lll}
\Upsilon_{abcd}=\Upsilon_{[ab][cd]}\,,\quad&\quad
\Upsilon_{[abc]d}=0\,,\quad&\quad\Upsilon_{e(ab)}{}^{e}=0\,,
\ea
\ee
such that in particular, $\Upsilon_{[abc]}{}^{a}=\frac{2}{3}\Upsilon_{a[bc]}{}^{a}=0$. Consequently  we get
\be
\Upsilon_{eab}{}^{e}=0\,.
\ee
The last condition~(\ref{lastcon}) then reduces to 
\be
\Upsilon_{[ab][cd]}+\Upsilon_{[cd][ab]} =0\,,
\ee
which further gives
\be
\Upsilon_{abcd}=-\Upsilon_{cdab}=\Upsilon_{dacb}+\Upsilon_{acdb}=-\Upsilon_{bcad}-\Upsilon_{acbd}=\Upsilon_{abcd}-2\Upsilon_{acbd}\,,
\ee
and hence,     the verification  of the uniqueness, 
\be
\Upsilon_{acbd}=0\,.
\ee
To summarize, the five conditions,  (\ref{compM}), (\ref{Gab}), (\ref{Gabc}), (\ref{Gcc}) and (\ref{lastcon}),  uniquely determines the connection (\ref{connectionGamma}).

\subsection{Semi-covariant derivative and  its complete covariantization} 
The infinitesimal diffeomorphic  transformation of the U-metric, 
\be
\delta_{X}M_{ab}=\hcL_{X}M_{ab}=\na_{ac}X_{b}{}^{c}+\na_{bc}X_{a}{}^{c}-\half M_{ab}\na_{cd}X^{cd}\,,
\label{deltaM}
\ee
induces upon the section condition,
\be
\delta_{X}(\partial_{ab}M_{cd})=\hcL_{X}(\partial_{ab}M_{cd})
+M_{ed}\partial_{ab}\partial_{cf}X^{ef}+M_{ce}\partial_{ab}\partial_{df}X^{ef}
-\half M_{cd}\partial_{ab}\partial_{ef}X^{ef}\,,
\ee
and hence
\be
\delta_{X}A_{abcd}=\hcL_{X}A_{abcd}-\half(\partial_{ab}\partial_{ce}X^{fe})M_{fd}-\half(\partial_{ab}\partial_{de}X^{fe})M_{fc}\,.
\label{deltaA}
\ee
It is  then straightforward to derive the variation  of the connection under diffeomorphism, 
\be
\delta_{X}\Gamma_{abc}{}^{d}=\hcL_{X}\Gamma_{abc}{}^{d}-\partial_{ab}\partial_{ce}X^{de}+\cP_{abc}{}^{dklm}{}_{n}\partial_{kl}\partial_{me}X^{ne}\,.
\label{deltaGamma}
\ee
For  consistency,  this expression is compatible  with  all the  properties  of the connection, such as
\be
\ba{ll}
\delta_{X}\Gamma_{(ab)c}{}^{d}=\hcL_{X}\Gamma_{(ab)c}{}^{d}\equiv0\,,
\quad&\quad
\delta_{X}\Gamma_{[abc]}{}^{d}=\hcL_{X}\Gamma_{[abc]}{}^{d}\equiv0\,,\\
\delta_{X}\Gamma_{c(ab)}{}^{c}=\hcL_{X}\Gamma_{cab}{}^{c}\equiv0\,,
\quad&\quad
\delta_{X}\left(\cP_{abcd}{}^{efgh}\Gamma_{efgh}\right)=\hcL_{X}\left(\cP_{abcd}{}^{efgh}\Gamma_{efgh}\right)\equiv0\,,
\ea
\ee
which can be easily verified using \textit{e.g.} the projection property `$\cP(1-\cP)=0$'   (\ref{projection2})  and  an identity,   
\be
\partial_{c(a}\partial_{b)d}X^{cd}=0\,.
\label{idX}
\ee
Further,  up to the section condition, we have
\be
\delta_{X}\Gamma_{abe}{}^{e}=\hcL_{X}\Gamma_{abe}{}^{e}-\partial_{ab}\partial_{ef}X^{ef}\,.
\label{deltatrace}
\ee
It is crucial to note that  the last term in (\ref{deltaGamma}), which we put hereafter\footnote{In the case of $N=5$,  $\,\Omega_{abcd}$ coincides with   `$\quarter H_{abcd}$'  in \cite{Park:2013gaj}.}
\be
\Omega_{abc}{}^{d}:=\cP_{abc}{}^{dklm}{}_{n}\partial_{kl}\partial_{me}X^{ne}\,,
\ee
generates    `anomalous' terms  in the variation of the semi-covariant derivative acting on  a generic   covariant tensor density,
\be
\ba{ll}
\delta_{X}(\na_{ab}T^{c_{1}c_{2}\cdots c_{p}}{}_{d_{1}d_{2}\cdots d_{q}})=&\hcL_{X}\left(\na_{ab}T^{c_{1}c_{2}\cdots c_{p}}{}_{d_{1}d_{2}\cdots d_{q}}\right)\\
{}&
-\sum_{i=1}^{p}T^{c_{1}\cdots e\cdots c_{p}}{}_{d_{1}d_{2}\cdots d_{q}}\Omega_{abe}{}^{c_{i}}
+\sum_{j=1}^{q}\Omega_{abd_{j}}{}^{e}T^{c_{1}c_{2}\cdots c_{p}}{}_{d_{1}\cdots e\cdots d_{q}}\,.
\ea
\label{semicovT}
\ee
The second line  is the  anomalous part. Hence, the semi-covariant derivative of a generic covariant tensor density is not necessarily covariant.\footnote{This is  also precisely analogous to DFT-geometry, \textit{c.f.} Eq.(20) of \cite{Jeon:2011cn},  where the anomalous part in the  diffeomorphic  variation of the DFT semi-covariant  derivative   is  dictated by a  six-index  projection operator.}  Nevertheless,  from (\ref{projection2}), (\ref{projectionsym}), (\ref{traceP}), (\ref{3aP}) and (\ref{idX}), $\Omega_{abcd}$ possesses  some nice properties,
\be
\ba{llll}
\Omega_{abcd}=\Omega_{[ab][cd]}=\Omega_{cdab}\,,\quad&\quad
\Omega_{[abc]d}=0\,,\quad&\quad\Omega_{acb}{}^{c}=0\,,\quad&\quad\Omega_{abcd}=\cP_{abcd}{}^{efgh}\Omega_{efgh}\,.
\ea
\label{Omegaprop}
\ee
These ensure that, for consistency, the followings  are exceptionally,  fully covariant.\\

\indent \textit{i)} The U-metric compatibility (\ref{compM}),
\be
\ba{ll}
\na_{ab}M_{cd}=0\,,
\quad&\quad
\delta_{X}(\na_{ab}M_{cd})=\hcL_{X}(\na_{ab}M_{cd})=0\,.
\ea
\label{exc1}
\ee
\indent \textit{ii)} Scalar density with an arbitrary extra weight,  
\be
\ba{ll}
\na_{ab}\phi=\partial_{ab}\phi+\half\omega\Gamma_{abc}{}^{c}\phi\,,\quad&\quad
\delta_{X}(\na_{ab}\phi)=\hcL_{X}(\na_{ab}\phi)\,.
\ea
\label{exc2}
\ee
\indent \textit{iii)} Kronecker delta symbol,
\be
\ba{ll}
\na_{ab}\delta^{c}_{~d}=0\,,\quad&\quad\delta_{X}(\na_{ab}\delta^{c}_{~d})=
\hcL_{X}(\na_{ab}\delta^{c}_{~d})=0\,.
\ea
\label{exc3}
\ee
In particular,  from (\ref{exc1}) and (\ref{exc2}), the integral measure, $|M|^{\frac{1}{4-N}}$ having  the extra weight, ${\omega=1}$, is also covariantly constant,
\be
\na_{ab}|M|^{\frac{1}{4-N}}=0\,.
\ee
~\\
\noindent The key characteristic   of  the semi-covariant derivative is that, by  (anti-)symmetrizing or contracting  the $\SLN$ vector  indices in an appropriate manner,    it can generate  completely  covariant derivatives  acting on a generic covariant tensor density, (\ref{fCOMPLETECOV}),
\begin{eqnarray}
&\na_{[ab}T_{c_{1}c_{2}\cdots c_{q}]}\,,&\label{COD1}\\
&\na_{ab}T^{a}\,,&\\
&\na^{a}{}_{b}T_{[ca]}+\na^{a}{}_{c}T_{[ba]}\,,&\label{Tasym}\\
&\na^{a}{}_{b}T_{(ca)}-\na^{a}{}_{c}T_{(ba)}\,,&\label{Tsym}\\
&~~\quad\quad\quad\na_{ab}T^{[abc_{1}c_{2}\cdots c_{q}]}~~~~~~~~\mbox{(divergence)}\,,&\label{COD2}\\
&\quad\quad\quad\na_{ab}\na^{[ab}T^{c_{1}c_{2}\cdots c_{q}]}~~~~~\mbox{(Laplacian)}\,.&
\label{COD3}
\end{eqnarray}
Note that the nontrivial values of $q$ in (\ref{COD1}), (\ref{COD2}) and  (\ref{COD3}) are restricted to $q=0,1,2,\cdots, N{-2}$ only, since the anti-symmetrization of more than $N$ number of $\SLN$ vector  indices is trivial. \\
\noindent Of course, from the U-metric compatibility, $\na_{ab}M_{cd}=0$,   the $\SLN$ vector  indices above can be freely  raised or lowered without spoiling  the  full covariance.  For example, the following  is also  fully covariant along with (\ref{COD1}),
 \be
 \na^{[ab}T^{c_{1}c_{2}\cdots c_{q}]}\,.
 \ee
Especially, for the case of $q=0$,  the divergence~(\ref{COD2}) reads explicitly,
\be
\na_{ab}T^{ab}=\partial_{ab}T^{ab}+\half(\omega-1)\Gamma_{abc}{}^{c}T^{ab}\,,
\ee
and hence,
\be
\na_{ab}T^{ab}=\partial_{ab}T^{ab}\quad\mbox{for}\quad\omega=1\,.
\ee
This is  a  useful relation for the discussion of  the    `total derivative' or  `surface integral' for  the      action.  \\

\noindent Successive  applications of the above procedure to a scalar and  a  vector  --or directly from (\ref{semicov2T})--   lead to  the following second-order  covariant derivatives,  
 \be
\ba{lll}
\na_{[ab}\na_{cd]}\phi=0\,,\quad&\quad\na_{[ab}\na_{cd}T_{e]}=0\,,\quad&\quad
\na_{[ab}\na_{c]d}T^{d}=0\,,
\ea
\ee
which turn out to be  all trivial  due to  (\ref{Gab}),  (\ref{Gabc}), (\ref{Gcc}) and    the section condition.  Similarly,  for arbitrary a scalar and  a vector, we have an identity, 
\be
\na_{[ab}\phi\,\na_{cd}T_{e]}=0\,.
\ee
It is worth while to note, from (\ref{semicovT}),  
\be
\ba{l}
(\delta_{X}-\hcL_{X})(\na_{ab}\na_{cd}T^{e_{1}e_{2}\cdots e_{p}}{}_{f_{1}f_{2}\cdots f_{q}})\\
{}=-\sum_{i=1}^{p}\left(T^{e_{1}\cdots g\cdots e_{p}}{}_{f_{1}\cdots f_{q}}\na_{ab}\Omega_{cdg}{}^{e_{i}}+\na_{ab}T^{e_{1}\cdots g\cdots e_{p}}{}_{f_{1}\cdots f_{q}}\Omega_{cdg}{}^{e_{i}}
+\na_{cd}T^{e_{1}\cdots g\cdots e_{p}}{}_{f_{1}\cdots f_{q}}\Omega_{abg}{}^{e_{i}}\right)\\
\quad~+\sum_{j=1}^{q}\left(\na_{ab}\Omega_{cdf_{j}}{}^{g}T^{e_{1}\cdots e_{p}}{}_{f_{1}\cdots g\cdots f_{q}}+\Omega_{abf_{j}}{}^{g}\na_{cd}T^{e_{1}\cdots e_{p}}{}_{f_{1}\cdots g\cdots f_{q}}+\Omega_{cdf_{j}}{}^{g}\na_{ab}T^{e_{1}\cdots e_{p}}{}_{f_{1}\cdots g\cdots f_{q}}\right)\\
\quad~+\Omega_{abc}{}^{g}\na_{gd}T^{e_{1}\cdots e_{p}}{}_{f_{1}\cdots f_{q}}
+\Omega_{abd}{}^{g}\na_{cg}T^{e_{1}\cdots e_{p}}{}_{f_{1}\cdots f_{q}}\,.
\ea
\label{semicov2T}
\ee
Further,  from (\ref{flastcon}) and (\ref{Omegaprop}), we have
\be
\Omega^{abcd}\Gamma_{abcd}=0\,,
\label{HG4}
\ee
which also implies with  (\ref{fGab}) and (\ref{Omegaprop}),
\be
\Omega^{abcd}\Gamma_{acbd}=0\,.
\label{HG42}
\ee

\subsection{Semi-covariant Riemann curvature and  its complete  covariantization\label{SECcurv}}
The commutator of the  semi-covariant derivative leads to  an expression,
\be
\ba{ll}
{}\left[\na_{ab},\na_{cd}\right]T^{e_{1}\cdots e_{p}}{}_{f_{1}\cdots f_{q}}=&
-\sum_{i=1}^{p}T^{e_{1}\cdots g\cdots e_{p}}{}_{f_{1}\cdots f_{q}}R_{abcdg}{}^{e_{i}}
+\sum_{j=1}^{q}R_{abcdf_{j}}{}^{g}T^{e_{1}\cdots e_{p}}{}_{f_{1}\cdots g\cdots f_{q}}\\
{}&\!\!\!\!\!\!\!\!\!\!\!+\left(2\Gamma_{ab[c}{}^{g}\delta_{d]}^{~h}-2\Gamma_{cd[a}{}^{g}\delta_{b]}^{~h}
-\half\Gamma_{abk}{}^{k}\delta_{c}^{~g}\delta_{d}^{~h}+\half\Gamma_{cdk}{}^{k}\delta_{a}^{~g}\delta_{b}^{~h}\right)\na_{gh}T^{e_{1}\cdots e_{p}}{}_{f_{1}\cdots f_{q}}\,,
\ea
\label{nana}
\ee
where $R_{abcde}{}^{f}$ denotes the  standard ``field strength" of the connection, 
\be
\ba{ll}
R_{abcde}{}^{f}&\!:=\partial_{ab}\Gamma_{cde}{}^{f}-\partial_{cd}\Gamma_{abe}{}^{f}+
\Gamma_{abe}{}^{g}\Gamma_{cdg}{}^{f}-\Gamma_{cde}{}^{g}\Gamma_{abg}{}^{f}\,,\\
{}&=\na_{ab}\Gamma_{cde}{}^{f}+\half\Gamma_{abg}{}^{g}\Gamma_{cde}{}^{f}+\Gamma_{cde}{}^{g}\Gamma_{abg}{}^{f}-\Gamma_{abc}{}^{g}\Gamma_{gde}{}^{f}-\Gamma_{abd}{}^{g}\Gamma_{cge}{}^{f}\,-\,\left[(a,b)\leftrightarrow(c,d)\right]\,,
\ea
\label{curvatureR}
\ee
which we call henceforth \textit{the fake curvature}. The fake curvature satisfies  identities that are rather trivial, 
\be
\ba{ll}R_{abcde}{}^{f}+R_{cdabe}{}^{f}=0\,,~~~&~~~R_{[abcd]e}{}^{f}=0\,.
\ea
\ee
On the other hand, from $\left[\na_{ab},\na_{cd}\right]M_{ef}=0$ for (\ref{nana}),   nontrivial identities are
\be
R_{abcdef}+R_{abcdfe}=0\,,
\label{efsym}
\ee
and hence,   we get\footnote{Eq.(\ref{Rsympro})  implies that  there exists essentially    \textit{only} one  fake `scalar'  curvature   one can construct   by contracting the indices of $R_{abcdef}$, which is   $R_{abc}{}^{abc}$~\cite{Park:2013gaj}.}
\be
R_{abcdef}=R_{[ab][cd][ef]}=-R_{[cd][ab][ef]}\,.
\label{Rsympro}
\ee
In particular,  the fake curvature is traceless, 
\be
R_{abcde}{}^{e}=0\,.
\ee
We  define the  \textit{semi-covariant  Riemann curvature},
\be
\ba{rl}
S_{abcd}:=&3\partial_{[ab}\Gamma_{e][cd]}{}^{e}+3\partial_{[cd}\Gamma_{e][ab]}{}^{e}+\quarter\Gamma_{abe}{}^{e}\Gamma_{cdf}{}^{f}+\half\Gamma_{abe}{}^{f}\Gamma_{cdf}{}^{e}\\
{}&+\Gamma_{ab[c}{}^{e}\Gamma_{d]ef}{}^{f}+\Gamma_{cd[a}{}^{e}\Gamma_{b]ef}{}^{f}+\Gamma_{ea[c}{}^{f}\Gamma_{d]fb}{}^{e}-\Gamma_{eb[c}{}^{f}\Gamma_{d]fa}{}^{e}\,.
\ea
\label{semicovS}
\ee
The semi-covariant  Riemann curvature can be  rewritten,  using the semi-covariant derivative,  
\be
\ba{rl}
S_{abcd}=&3\na_{[ab}\Gamma_{e][cd]}{}^{e}+3\na_{[cd}\Gamma_{e][ab]}{}^{e}-\quarter\Gamma_{abe}{}^{e}\Gamma_{cdf}{}^{f}-\half\Gamma_{abe}{}^{f}\Gamma_{cdf}{}^{e}\\
{}&-\Gamma_{ab[c}{}^{e}\Gamma_{d]ef}{}^{f}-\Gamma_{cd[a}{}^{e}\Gamma_{b]ef}{}^{f}
-\Gamma_{ea[c}{}^{f}\Gamma_{d]fb}{}^{e}+\Gamma_{eb[c}{}^{f}\Gamma_{d]fa}{}^{e}\,,
\ea
\ee
or in terms of the fake curvature,
\be
\ba{rl}
S_{abcd}=&R_{abe[cd]}{}^{e}+R_{cde[ab]}{}^{e}
-\half\Gamma_{abe}{}^{f}\Gamma_{cdf}{}^{e}
+\quarter\Gamma_{abe}{}^{e}\Gamma_{cdf}{}^{f}\\
{}&
+\half\Gamma_{ead}{}^{f}\Gamma_{fcb}{}^{e}
+\half\Gamma_{eda}{}^{f}\Gamma_{fbc}{}^{e}
-\half\Gamma_{ebd}{}^{f}\Gamma_{fca}{}^{e}
-\half\Gamma_{edb}{}^{f}\Gamma_{fac}{}^{e}\\
{}&
+\quarter\Gamma_{eaf}{}^{f}\Gamma_{cdb}{}^{e}
+\quarter\Gamma_{ecf}{}^{f}\Gamma_{abd}{}^{e}
-\quarter\Gamma_{ebf}{}^{f}\Gamma_{cda}{}^{e}
-\quarter\Gamma_{edf}{}^{f}\Gamma_{abc}{}^{e}\,.
\ea
\ee
By construction, it satisfies symmetric properties,
\be
\ba{ll}
S_{abcd}=S_{[ab][cd]}\,,\quad&\quad S_{abcd}=S_{cdab}\,,
\ea
\ee
and thanks to  the section condition, it meets  the Bianchi identity,
\be
S_{[abc]d}=0\,.
\ee
Under arbitrary variation of  the connection, $\delta\Gamma_{abc}{}^{d}$,  which is,   from (\ref{fGab}), (\ref{fGabc}), (\ref{fGcc}),  subject to
\be
\ba{lll}
\delta\Gamma_{(ab)c}{}^{d}=0\,,\quad&\quad
\delta\Gamma_{[abc]}{}^{d}=0\,,\quad&\quad
\delta\Gamma_{c(ab)}{}^{c}=0\,,
\ea
\ee
the fake curvature  transforms in a somewhat complicated manner, 
\be
\delta R_{abcde}{}^{f}=\na_{ab}\delta\Gamma_{cde}{}^{f}+\half\Gamma_{abg}{}^{g}\delta\Gamma_{cde}{}^{f}
-\Gamma_{abc}{}^{g}\delta\Gamma_{gde}{}^{f}
-\Gamma_{abd}{}^{g}\delta\Gamma_{cge}{}^{f}\,-\,\left[(a,b)\leftrightarrow(c,d)\right]\,.
\label{deltaR6}
\ee
On the other hand, \textit{the semi-covariant  Riemann curvature transforms as   total derivative},
\be
\delta S_{abcd}=3\na_{[ab}\delta\Gamma_{e][cd]}{}^{e}+
3\na_{[cd}\delta\Gamma_{e][ab]}{}^{e}\,.
\label{deltaS4}
\ee
In fact, this is  the crucial `defining'  property of the semi-covariant  Riemann curvature which    we prerequired   to  derive   the  expression~(\ref{semicovS}). \\

\noindent Especially, under  diffeomorphism (\ref{deltaGamma}), while the connection  changes, 
\be
\ba{ll}
\delta_{X}\Gamma_{abc}{}^{d}=\hcL_{X}\Gamma_{abc}{}^{d}-\partial_{ab}\partial_{ce}X^{de}+\Omega_{abc}{}^{d}\,,\quad&\quad
\Omega_{abcd}=\cP_{abcd}{}^{klm}{}_{n}\partial_{kl}\partial_{me}X^{ne}\,,
\ea
\label{Rvar1}
\ee
the fake curvature varies, 
\be
\ba{ll}
\delta_{X}R_{abcdef}-\hcL_{X}R_{abcdef}=&\na_{ab}\Omega_{cdef}+
\half\Gamma_{abg}{}^{g}\Omega_{cdef}-\Gamma_{abc}{}^{g}\Omega_{gdef}-\Gamma_{abd}{}^{g}\Omega_{cgef}\\
{}&+\partial_{ab}\partial_{ch}X^{gh}\Gamma_{gd[ef]}+\partial_{ab}\partial_{dh}X^{gh}\Gamma_{cg[ef]}
-\half\partial_{ab}\partial_{gh}X^{gh}\Gamma_{cd[ef]}\\
{}&\,-\,\left[(a,b)\leftrightarrow(c,d)\right]\,,
\ea
\label{varR}
\ee
and the semi-covariant  Riemann curvature transforms  \textit{neatly}, 
\be
\delta_{X}S_{abcd}=\hcL_{X}S_{abcd}
+2\na_{e[a}\Omega_{b][cd]}{}^{e}
+2\na_{e[c}\Omega_{d][ab]}{}^{e}\,.
\ee
Like the semi-covariant derivative~(\ref{semicovT}), the anomalous terms  are   dictated  by the projection operator.\footnote{Again, this is precisely analogous to the DFT-geometry, \textit{c.f.} Eq.(27) in Ref.\cite{Jeon:2011cn}.} 
Therefore, as the name indicates, the fake curvature, $R_{abcdef}$,  is not  covariant. Yet,  with the nice properties of $\Omega_{abcd}$  (\ref{Omegaprop}),  the semi-covariant  Riemann curvature can be completely  covariantized,  such as   Ricci and scalar curvatures:\footnote{Note that $S_{ab}$ and $S$ are  related to `$\cR_{ab}$' and `$\cR$' in \cite{Park:2013gaj} by   factor two:  $S_{ab}=2\cR_{ab}$, $\,S=2\cR$.}
\be
\ba{ll}
S_{ab}:=S_{acb}{}^{c}=S_{ba}\,,\quad&\quad
\delta_{X}S_{ab}=\hcL_{X}S_{ab}\,,\\
S:=M^{ab}S_{ab}=S_{ab}{}^{ab}\,,\quad&\quad
\delta_{X}S=\hcL_{X}S=\half X^{ab}\partial_{ab}S\,.
\ea
\label{RicciScalar}
\ee
For later use, it is worth while to have  an explicit  expression of the completely covariant  scalar curvature,
\be
S=-2\partial_{ab}(2A^{cab}{}_{c}+A^{abc}{}_{c})+A_{abcd}A^{abcd}-4A_{abcd}A^{acbd}-\textstyle{\frac{1}{2}} A_{abc}{}^{c}A^{abd}{}_{d}-4A_{cab}{}^{c}A^{abd}{}_{d}+4A_{cab}{}^{c}A^{dba}{}_{d}\,,
\label{ScalarA}
\ee
where, as defined before (\ref{defA}),
\be
A_{abcd}=-\half\partial_{ab}M_{cd}+\textstyle{\frac{1}{2(N-4)}}M_{cd}\partial_{ab}\ln\left|M\right|\,.
\ee


\subsection{Action and  the Einstein equation of motion}
From  (\ref{deltaS4}),  it is straightforward to  derive the  variation    of  the fully covariant scalar curvature,    
\be
\delta S=2\delta M^{ab}S_{ab}+6\na_{[ab}\left(\delta\Gamma_{e]cd}{}^{e}M^{ac}M^{bd}\right)
\,.
\label{deltaS0}
\ee
Hence, disregarding  surface integral,  arbitrary variation of the U-metric induces the following  transformation  of the U-gravity   action~(\ref{factionR}), 
\be
\delta\left(\dis{\int_{\Sigma}}M^{\frac{1}{4-N}}S\right)=
\dis{\int_{\Sigma}}M^{\frac{1}{4-N}}\delta M^{ab}\left(2S_{ab}+\textstyle{\frac{1}{N-4}}M_{ab}S\right)\,,
\ee
which leads to the \textit{Einstein equation of motion}~(\ref{fEOM}). Further, from the invariance of the action under   diffeomorphism~(\ref{deltaM}), a  \textit{conservation relation}~(\ref{fconservation})  follows.\footnote{The   \textit{conservation relation}~(\ref{fconservation})  may be also directly verified using the Jacobiator of the semi-covariant derivative, \textit{c.f.~}Eq.(4.3) in Ref.\cite{Park:2013gaj}.}

\subsection{Reductions\label{SECREDUCTION}}  
Here we discuss the reduction of $\SLN$ U-gravity  upon  each section, $\Sigma_{N-1}$ and $\Sigma_{3}$ separately. The resulting gravitational  actions contain  $(N{-2})$-form or two-form  potentials as well as scalars.\footnote{Hence, \textit{a priori}, the $\SLN$ duality differs from  the Ehlers group of the  Einstein-Hilbert pure gravity action, \textit{c.f.}~\cite{Hohm:2013jma}.  }
\begin{enumerate}
\item  \textit{Reduction upon $\Sigma_{N-1}$.}\\
In order to perform the Riemannian reduction to the  $(N-1)$-dimensional section, $\Sigma_{N-1}$~(\ref{fMsection}), we parametrize the U-metric by~\cite{Berman:2010is,Park:2013gaj}
\be
\ba{ll}
 M_{ab} = 
 \left( \begin{array}{cc}
 \frac{g_{\alpha\beta}}{\sqrt{|g|}} & v_{\alpha} \\
		  v_{\beta} & \sqrt{|g|} \left(- e^{\phi} + v^{2}\right)
                 \end{array} \right) \,,&M^{ab}=\left(\ba{cc}
\sqrt{|g|}(g^{\alpha\beta}-e^{-\phi}v^{\alpha}v^{\beta})~&~~e^{-\phi}v^{\alpha}\\
e^{-\phi}v^{\beta}~&~-\frac{e^{-\phi}}{\sqrt{|g|}}\ea\right)\,.
\ea
\label{PARAN}
\ee
Here $\phi$, $v^{\alpha}$ and $g_{\alpha\beta}$ denote a scalar, a vector and  a Riemannian   metric  on $\Sigma_{N-1}$, such that $v_{\alpha}=g_{\alpha\beta}v^{\beta}$, $v^{2}=g^{\alpha\beta}v_{\alpha}v_{\beta}$  and $g=\det(g_{\alpha\beta})$.  
The vector can be freely  dualized   to an $(N{-2})$-form potential which  may couple to an  $(N{-3})$-brane.

With the Riemannian ansatz~(\ref{PARAN}), the U-gravity scalar curvature~(\ref{ScalarA})    reduces  to (\ref{fREDUCEDN}) which agrees with \cite{Park:2013gaj} when $N=5$.   Consistently,  the generalized Lie derivative~(\ref{fgLNw}) decomposes into  the $(N{-1})$-dimensional  ordinary  Lie derivative and the  gauge symmetry of the $(N{-2})$-form potential. We refer the readers to  Eq.(5.6) of \cite{Park:2013gaj} for the explicit demonstration in the case of $\SLf$.

It is crucial  to note that   a nontrivial   assumption has been  implicitly made to write  the ansatz~(\ref{PARAN}),  namely that the upper left $(N-1)\times(N-1)$ block of the U-metric is \textit{non-degenerate} and hence we are allowed to  write  ``\,${g_{\alpha\beta}}/{\sqrt{|g|}}$\," there.  However,   the rank of the  $(N-1)\times(N-1)$ block  can be     $N-2$ (but not less than that  for the  U-metric to be non-degenerate). The  degenerate case   then corresponds  to a non-Riemannian background where the Riemannian metric ceases to exist.  Nevertheless,  $\SLN$ U-gravity has no problem with  that.  One example of such a non-Riemannian background  is given by a U-metric of which the only nontrivial components are $M_{1N}=M_{N1}$ and $M_{\hat{\alpha}\hat{\beta}}$ with $\,2\leq\hat{\alpha},\hat{\beta}\leq N-1$.

\item   \textit{Reduction upon $\Sigma_{3}$.}\\
For the Riemannian reduction of U-gravity to the  three-dimensional section, $\Sigma_{3}$~(\ref{f3section}),  we  put~\cite{Blair:2013gqa},
\be
\ba{ll}
 M_{ab} =\!\left(\!\ba{cc}
\sqrt{|\tilde{g}|}(\tilde{g}_{\mu\nu} +
e^{\tphi} \tilde{v}^k{}_{\mu} \tilde{v}_{k\nu}) & e^{\tphi}\tilde{v}_{j\mu} \\
e^{\tphi} \tilde{v}_{i\nu} & \frac{e^{\tphi}}{\sqrt{|\tilde{g}|}} \tilde{\cM}_{ij}
\ea\!\right)\,,&
M^{ab}=\!\left(\!\ba{cc}
\frac{\tg^{\mu\nu}}{\sqrt{|\tg|}}~&~
-\tv^{j\mu}
\\
-\tv^{i\nu}~&~\sqrt{|\tg|}(e^{-\tphi}\tilde{\cM}^{ij}+\tv^{i\lambda}\tv^{j}{}_{\lambda})
\ea\!
\right)\,.
\ea
\label{PARA3} 
\ee
Here, to be consistent with the `lower' index of the dual coordinates, $\tilde{x}_{\mu}$, the Riemannian  metric is $\tilde{g}^{\mu \nu}$ having  `upper' indices,  with the determinant, $\tilde{g}\equiv\det(\tg^{\mu \nu})$;   $\tilde{\cM}_{ij}$ is a symmetric $(N-3)\times (N-3)$ unit determinant matrix; and $\tv_{i\mu}$ are $(N-3)$ copies of vectors while  $\tv^{i\mu}=\tilde{\cM}^{ij}\tg^{\mu\nu}\tv_{j\nu}$.  The vectors can be dualized to two-form potentials. 

With  the Riemannian ansatz~(\ref{PARA3}), the U-gravity scalar curvature~(\ref{ScalarA})  reduces  to (\ref{fREDUCED3}) which features {$\mathbf{SL}(N-3)$ {S-duality}} and agrees with \cite{Blair:2013gqa} when  $N=5$.  Consistently,  the generalized Lie derivative~(\ref{fgLNw}) decomposes into  the three-dimensional  ordinary  Lie derivative and the  gauge symmetry of the two-form potentials.  We refer the readers to  Eq.(3.8) of  \cite{Blair:2013gqa}  for the explicit demonstration in the case of $\SLf$.

Writing (\ref{PARA3}), it has been   assumed  that the upper left $3\times 3$ block of $M^{ab}$ is non-degenerate.  But, in general,  its rank can be less than $3$. In fact, when $N\geq 6$ the whole block can vanish: for example  the  only nontrivial components of the inverse of the U-metric can be, $M^{\mu\hat{\imath}}=M^{\hat{\imath}\mu}$ and $M^{\tilde{\imath}\tilde{\jmath}}$ where $\hat{\imath}=4,5,6$ and $7\leq\tilde{\imath},\tilde{\jmath}\leq N$. When $N=5$, the rank of the $3\times 3$ block is either  $3$ (non-degenerate)   or   at least $2$ (degenerate).

\end{enumerate}

\section{Outlook\label{SECCOMMENTS}}
On the extended-yet-gauged spacetime, the usual   differential one-form of the coordinate,  $\rmd x^{ab}$, is not  invariant under the coordinate gauge symmetry~(\ref{fCGS}), and thus  needs to be  \textit{gauged},  \textit{c.f.~}\cite{Park:2013gaj}
\be
\ba{ll}
Dx^{ab}:=\rmd x^{ab}-A^{ab}\,,\quad&\quad A^{ab}\partial_{ab}\equiv 0\,.
\ea
\ee
Here a  connection has been introduced which assumes the same `value' as the  coordinate gauge symmetry generator, or the shift~(\ref{fTensorCGS}).  Essentially it gauges away the orthogonal directions to a chosen section.  The gauged  one-form can be then  used to construct an $\SLN$ duality manifest  world-volume action for  an  $(N-3)$-brane  propagating   in the  extended-yet-gauged spacetime, as done for a string in \cite{Park:2013gaj} (\textit{c.f.}  \cite{Hohm:2013jaa,Blair:2013noa,Hatsuda:2012vm}).\\

\noindent  The notion of the cosmological `constant' depends on the kind of  differential geometry in use~\cite{Jeon:2011cn}.  In $\SLN$ U-gravity, the natural cosmological constant term reads 
\be
\dis{\int_{\Sigma}}M^{\frac{1}{4-N}}\Lambda\,.
\ee
Yet, from the Riemannian point of view,   \textit{i.e.~}(\ref{fPARAN}) or (\ref{fPARA3}), this term corresponds  to an exponential potential of  the  scalar.  This might provide a new spin on the cosmological constant problem, \textit{c.f.}~\cite{Kan:2011vg,Wu:2013sha,Wu:2013ixa}.\\

\noindent Recent studies  indicate that, in order to identify the DFT/EFT origins   of all the known  lower dimensional gauged supergravities, it is necessary  to `relax'  the section condition~\cite{Geissbuhler:2011mx,Grana:2012rr,Dibitetto:2012rk,Berman:2012uy,Musaev:2013rq,Aldazabal:2013mya,Berman:2013uda,Geissbuhler:2013uka,Aldazabal:2013sca,Berman:2013cli}.    The geometric insight of the  extended-yet-gauged spacetime is then  somewhat  unclear.   Perhaps, the strict invariance under the coordinate gauge symmetry~(\ref{fTensorCGS}) may  not be  the only way to realize the    extended-yet-gauged spacetime.  The final  geometric understanding    is incomplete. In this line,  it is worth while to   note  an interesting recent development~\cite{Cederwall:2014kxa} where the flat $\ODD$ metric in DFT is  promoted  to a generic curved one and  the section condition is accordingly modified. \\

\noindent Understanding of  the global and topological   aspects  of $\SLN$ U-gravity  is desirable along with   further geometric insights into the   non-Riemannian backgrounds, \textit{c.f.~}\cite{Hull:2006qs,Hohm:2012gk,Park:2013mpa,Hohm:2013nja,Berman:2014jba,Cederwall:2014kxa,Papadopoulos:2014mxa,Lee:2014mla}.  \\

\noindent  Taking  $N=11$,   $\mathbf{SL}(11)$ U-gravity  may provide an $\mathbf{SL}(11)$ U-duality manifest reformulation of the ten-dimensional massive  type IIA supergravity~\cite{Romans:1985tz} with the identification of the ten-form flux as the cosmological constant~\cite{Polchinski:1995mt}. This will be  in analogous to the $\Ott$ T-duality    manifest unification   of  IIA and IIB supergravities~\cite{Jeon:2012hp} (\textit{c.f.}~\cite{Hohm:2011zr,Hohm:2011dv}). \\

\noindent   In view of the Dynkin diagram (Table~\ref{TableDynkin}),   putting    $\mathbf{SL}(11)$  U-gravity and $\Ott$ DFT together,     one may  anticipate the whole  $E_{11}$ structure to emerge. A tantalizing clue comes from the   RR nine-form potential, which is dual to the vector in the $\Sigma_{10}$ parametrization of the U-metric~(\ref{fPARAN}).  In  the $\cN=2$ $D=10$ SDFT of \cite{Jeon:2012hp}, the local Lorentz group is doubled  to be $\Spin(1,9)_{L}\times\Spin(1,9)_{R}$  and the whole RR-sector is represented by a  single $\Spin(1,9)_{L}\times\Spin(1,9)_{R}$ bi-spinorial object which is \textit{a priori} $\Ott$ \textit{singlet}.  After diagonal gauge fixing of the doubled local  Lorentz group,   the single bi-spinorial object may decompose into various RR $p$-form potentials which are  no longer $\Ott$ singlet but form   an $\Ott$ spinor,  to agree with   \cite{Hohm:2011zr,Hohm:2011dv}.  On the other hand, in  $\mathbf{SL}(11)$ U-gravity,   the $\mathbf{SL}(11)$ group  does not mix the  RR nine-form potential with other  RR $p$-form potentials, since only the  nine-form potential  enters the parametrization of the  U-metric~(\ref{fPARAN}).   This might shed light on  the $E_{11}$ duality manifest reformulation of the maximal supergravity. But here we can only speculate.  \\

\section*{Acknowledgements} 
We would like to thank 	Chris D. A. Blair, Jose Juan Fernandez-Melgarej,  Sung Moon Ko,    Charles Melby-Thompson, Rene Meyer  and especially Henning Samtleben  for various helpful  discussions.   This work was  supported by Basic Science Research Program through the National Research Foundation of Korea (NRF)   with the Grant    No. 2012R1A2A2A02046739 and   No.
2013R1A1A1A05005747.\\



\end{document}